%% file: main.tex
\documentclass[fleqn,usenatbib]{mnras}
\usepackage{newtxtext,newtxmath}
\usepackage[T1]{fontenc}
\usepackage{ae,aecompl}

\usepackage{graphicx}	% Including figure files
\usepackage{amsmath}	% Advanced maths commands
\usepackage{amssymb}	% Extra maths symbols
\usepackage{adjustbox}
\usepackage{comment}
\usepackage{siunitx}
\usepackage{float}
\usepackage[table,xcdraw]{xcolor}
\usepackage{xspace}
\defcitealias{h18}{H18}
\defcitealias{f18}{F18}
\defcitealias{ma18}{M18}
\defcitealias{s17}{S17}
\defcitealias{w09data}{W09}
\defcitealias{bs12}{BS12}
\defcitealias{j95}{J95}
\defcitealias{b14}{B14}
\newcommand {\JP} {\citetalias{j95}\xspace}
\newcommand {\BP} {\citetalias{b14}\xspace}
\newcommand {\HP} {\citetalias{h18}\xspace}
\newcommand {\FP} {\citetalias{f18}\xspace}
\newcommand {\MP} {\citetalias{ma18}\xspace}
\newcommand {\SP} {\citetalias{s17}\xspace}
\newcommand {\WA} {\citetalias{w09data}\xspace}
\newcommand {\BS} {\citetalias{bs12}\xspace}

\newcommand{\te}[1]{\text{#1}}
\newcommand {\MW} {Milky Way }
\newcommand {\kms} {\,{\rm km\,s}^{-1}}
\newcommand {\pc} {\,{\rm pc}}
\newcommand {\kpc} {\,{\rm kpc}}

\newcommand {\Gyr} {\,{\rm Gyr}}

\newcommand {\lsun} {\,{\rm L}_\odot}
\newcommand {\msun} {\,{\rm M}_\odot}
\newcommand {\vlos} {\,{V}_{\rm los}}
\newcommand {\mvir} {\,{M}_{200}}
\newcommand {\rc} {\,{r}_{\rm c}}
\newcommand {\vsys} {\,{V}_{\rm sys}}
\newcommand {\rapo} {\,{r}_{\rm apo}}
\newcommand {\rperi} {\,{r}_{\rm peri}}
\newcommand{\vir}[1]{``#1''}
\newcommand{\red}{\color{red}}

\newcommand{\gaia}{\textit{Gaia} }

\usepackage{etoolbox}
\makeatletter
\patchcmd\@combinedblfloats{\box\@outputbox}{\unvbox\@outputbox}{}{\errmessage{\noexpand patch failed}}
\makeatother
\title[The effect of tides on Sculptor]{The effect of tides on the Sculptor dwarf spheroidal galaxy}

\author[G. Iorio et al.]{
G. Iorio$^{1}$\thanks{giorio@ast.cam.ac.uk},
C. Nipoti$^{2}$,
G. Battaglia$^{3,4}$,
A. Sollima$^{5}$
\\
$^{1}$Institute of Astronomy, University of Cambridge, Madingley Road, Cambridge CB3 0HA, UK\\
$^{2}$Dipartimento di Fisica e Astronomia, Universit\`a di Bologna, via Gobetti 93/2, I-40129, Bologna, Italy\\
$^{3}$Instituto de Astrof\'{i}sica de Canarias (IAC),
C/V\'{i}a L\'{a}ctea, s/n, 38205, San Crist\'{o}bal de la Laguna, Tenerife, Spain\\
$^{4}$Departamento de Astrof\'{i}sica, Universidad de La Laguna, 38206, San Crist\'{o}bal de la Laguna, Tenerife, Spain\\
$^{5}$INAF - Osservatorio di astrofisica e scienza dello spazio di Bologna, via Gobetti 93/3, 40129 Bologna, Italy\\
}
\date{Accepted XXX. Received YYY; in original form ZZZ}

\pubyear{2019}

\begin{document}
\label{firstpage}
\pagerange{\pageref{firstpage}--\pageref{lastpage}}
\maketitle

\begin{abstract}

\input{include/abstract}
\end{abstract}


%%%%%%%%%%%%%%%%%%%%%%%%%%%%%%%%%%%%%%%%%%%%%%%%%%

%%%%%%%%%%%%%%%%% BODY OF PAPER %%%%%%%%%%%%%%%%%%

\input{include/Introduction}
\input{include/Sculptor/Sculptor}

\input{include/Orbital/Orbital}
\input{include/Dynamics/DynamicalModel}

\input{include/Simulations/Simulations}

\input{include/Results/Results}
\input{include/Discussion/Discussion}

\section*{Acknowledgements}
The authors thank Raffaele Pascale for useful discussions and Luis Cicu\'endez for the analysis of the light distribution of Sculptor.  
We thank the anonymous referee for suggestions that helped to improve this manuscript.
In this work we made use of the  \texttt{Python} modules \texttt{matplotlib} \citep{matplotlib}, \texttt{numpy} and \texttt{scipy} \citep{scipy}. We also made use of the comology calculator by \cite{CosmoCalc}.
G.I.\ is supported by the Royal Society Newton  International Fellowship (NF170902). 
G.B. gratefully acknowledges financial support from Spanish Ministry of Economy and Competitivness 
(MINECO) under the Ramon y Cajal Programme (RYC-2012-11537) and the grant AYA2017-89076-P.

%%%%%%%%%%%%%%%%%%%%%%%%%%%%%%%%%%%%%%%%%%%%%%%%%%



\bibliographystyle{mnras}
\bibliography{main} % if your bibtex file is called example.bib

%%%%%%%%%%%%%%%%%%%%%%%%%%%%%%%%%%%%%%%%%%%%%%%%%%

%%%%%%%%%%%%%%%%% APPENDICES %%%%%%%%%%%%%%%%%%%%%
%\input{include/appendix/appendix}
\appendix
\input{include/appendix/mockobs}

%%%%%%%%%%%%%%%%%%%%%%%%%%%%%%%%%%%%%%%%%%%%%%%%%%

% Don't change these lines
\bsp	% typesetting comment
\label{lastpage}
\end{document}

%% file: include/abstract.tex
Dwarf spheroidal galaxies (dSphs)  appear to be some of the most dark matter dominated objects in the Universe. Their dynamical masses are commonly derived using the kinematics of stars  under the assumption of equilibrium. However, these objects are satellites of massive galaxies (e.g.\ the Milky Way) and thus can be influenced by their tidal fields.
We investigate the implication of the assumption of equilibrium focusing on the Sculptor dSph by means of ad-hoc $N$-body simulations tuned to reproduce the observed properties of Sculptor following the evolution along some observationally motivated orbits in the Milky Way gravitational field. For this purpose, we used state-of-the-art spectroscopic and photometric samples of Sculptor's stars. 
We found that the stellar component of the simulated object is not directly influenced by the tidal field, while $\approx 30\%-60\%$ the mass  of the more diffuse DM halo is stripped. 
We conclude that, considering the most recent estimate of the Sculptor proper motion, the system is not affected by the tides and the stellar kinematics represents a robust tracer of the internal dynamics. In the simulations that match the observed properties of Sculptor, the present-day dark-to-luminous mass ratio is $\approx 6$ within the stellar half-light radius ($\approx0.3$ kpc) and $>50$ within the maximum radius of the analysed dataset ($\approx1.5^\circ\approx2$ kpc).
%MDM_toMs=5.7 at Rh  and 46.53 at Rstar (EH18_13)
%MDM_toMs=5.9 at Rh  and 50.69 at Rstar (FH18_10)

%% file: include/Introduction.tex
\section{Introduction}

Dwarf galaxies are the most numerous galaxies in the Universe \citep{Ferguson94}: around the Local Group, about 70\% of galaxies are dwarfs \citep{mc}.  The fraction remains this high also considering the Local Volume within 11 Mpc from the Milky Way \citep{dale09}.
Understanding their structure, formation and evolution is a fundamental goal for astrophysics and cosmology \citep{klypin99, moore99, bullock00, grebel03, boylan11, Mayer11, Ander13, battaglia13, walker13, readiorio17}.
Among them, the dwarf spheroidal galaxies (dSphs), satellites of the Milky Way, 
are close enough to measure the kinematics of individual stars \citep[e.g.\ ][]{kleyna02, tol04, battagliasculp, munoz06, koch09, w09data,ma18}.  
The analysis of kinematic dataset with equilibrium models leads to the finding that these galaxies are some of the most dark matter (DM) dominated object known to date \citep{mateo, gilmore07}, with dynamical mass-to-light ratios up  to several 100s  (e.g.\ \citealt{mateo93,munoz06b,battagliasculp,koposov11,battaglia11}; see \citealt{mateo} and \citealt{battaglia13} for further information on the mass-to-light ratios of dSphs).
Therefore, these objects are ideal laboratories to test theories on DM physics and cosmology \citep{nfw, readcore, maccio10,walker13,kennedy14,kubik17}. 

However, since these dSphs are orbiting in the gravitational field of the MW, it is possible that the stellar tracers used to infer the dynamical state of these objects are influenced by tides and the assumption of equilibrium could introduce severe biases in the mass estimate \citep{kroupa97,fleck03, metz07,Dabringhausen16}. For example, stars in undetected   tidal tails  (too faint or along the line-of-sight) could  \vir{pollute} the kinematic samples inflating the velocity dispersion, with a consequent overestimate the DM content, if inferred using equilibrium dynamical models \citep[see e.g.][]{klimentowski09,read06tidal}.

Several works have focused on the effects of the  host gravitational field  on the internal properties of dwarf galaxies by means of $N$-body simulations \citep[e.g.][]{read06tidal,penarrubia08,penarrubia09,frings17,fattahi18}. The effect of tides on these galaxies strongly depends on their orbital history and the degree of pollution of tidally perturbed stars depends on the relative orientation of the  internal part of the tides with respect to the line of sight.
Therefore, the natural step forward is to perform similar analysis focusing on the properties of specific dSphs orbiting around the Milky Way. These works are based on $N$-body simulations tuned to reproduce the observed properties of the analysed dSph after their orbital evolution in the Milky Way tidal field. 
Works of this kind have been applied to Carina \citep{UralCarina}, Leo I \citep{sohn07,mateo08}, Fornax \citep{Fornax15} and to the recently discovered Crater II \citep{SandersCrater}. However, one of the biggest source of uncertainty is the estimate of the proper motions that are indispensable  to put the objects in  realistic orbits to measure the actual effect of tides on the observables. The choice of a well-motivated orbit is essential to consider the right amount of tidal disturbance suffered by the dSphs and to rightly evaluate the degree of tidally perturbed stars contaminating the observed dataset. Luckily, at the moment we are living in the golden era of proper motion measurements thanks to the revolutionary  Gaia DR2 catalogue \citep{GaiaDR2,h18,f18} and the very long baseline   supplied by the HST measurements \citep[e.g.][]{s17}.

Following  the work of \cite{Fornax15}, focused on the Fornax dwarf spheroidal, the aim of this paper is to investigate and quantify the effects of the tides on the structure and kinematics of the Sculptor dwarf spheroidal. 
\cite{Fornax15} found that in the case of Fornax  tides have an almost negligible effect on the observed structure and kinematics of the stellar component, even considering the most eccentric orbits compatible with the observational constraints. 
Similarly to Fornax, the Sculptor dSph is one of the best studied \MW satellites  
\citep[see e.g.][]{battagliasculp,penarrubia09,w09,Breddels14,tolstoy14,strigariSculp,ma18}, so  we can rely on high-quality spectroscopic data against which to compare our models \citep{w09data,bs12}.
Moreover, the predicted Sculptor orbit is less external (smaller pericentre) than that of Fornax, so we expect that it could be more influenced by tides. 
For this aim, we ran a set of $N$-body simulations following the evolution of the structural and kinematic properties of Sculptor-like objects on orbits around the Milky Way that are consistent with the most recent observational constraints. Then, simulating \vir{observations} in the last snapshot of the simulations, we investigated whether the observed kinematics of the stars is a robust tracer for the current dynamical state of Sculptor or the tides are introducing non-negligible biases.

The paper is organised as follows. 
In Sec.\ \ref{sec:theScl} we summarise  Sculptor's observational properties and we present a clean sample of Sculptor's member stars that we use to analyse the kinematics and dynamics of Sculptor.
In Sec.\ \ref{sec:orbital} we determine the possible Sculptor's orbits around the \MW making use of the most recent and accurate Sculptor's proper motion estimates.
In Sec.\ \ref{sec:dynmod}, we use the information obtained in the previous sections to define a  dynamical model of Sculptor compatible with the observable properties.
Then, in Sec.\ \ref{sec:Nbodysim} we describe the set-up of $N$-body simulations, whose results are presented in  Sec.\ \ref{sec:fresults}. Finally, 
in Sec.\ \ref{sec:dicussion} we summarise and discuss the main results of this work.

%% file: include/Sculptor/Sculptor.tex
\section{Observed properties of the Sculptor dSph} \label{sec:theScl}

\input{include/Sculptor/SculptorInfo}

\input{include/Sculptor/VdispProfile}

%% file: include/Sculptor/SculptorInfo.tex
\subsection{Structure} \label{sec:obs}

\begin{table}
\centering
\begin{tabular}{ccc}
\hline
Parameter            & Value  & Reference \\
\hline
($\ell$, $b$)             & ($287.53^\circ$, $-83.16^\circ$)    & 1 \\
$D$                  & $86\pm6^\dagger$ kpc                        & 2 \\
$L_*$         & $2.3\times10^6 \  \lsun$     & 1 \\
$\Upsilon_*$       & $2$                                 & 3 \\
$R_\te{h}$ & $11.30\arcmin\pm1.60\arcmin$ ($283\pm45 \ \pc^\dagger$)        & 1 \\
\hline
$b_*$ & $11.15\arcmin\pm0.18\arcmin$ ($279\pm4.5 \ \pc^\dagger$)        & C \\
$\epsilon$ & $0.226\pm0.014$         & C \\
$PA$ & $97.4^\circ\pm1.2^\circ$       & C \\
\hline
$V_\te{sys}$  & $111.6\pm0.2 \ \kms$ & T \\
$\sigma_\te{los}(R<R_\te{h})$  & $8.4\pm0.2 \ \kms$ & T \\
$\sigma_\te{los}(R<1.5^\circ)$  & $8.9\pm0.2 \ \kms$ & T \\ \hline
\end{tabular}
\caption{Sculptor's observational parameters. From the top to the bottom: Galactic coordinates ($\ell$,$b$), distance $D$, total luminosity in the $V$ band $L_*$, stellar mass-to-light ratio in the $V$ band $\Upsilon_*$,  half-light radius $R_\te{h}$, best-fit Plummer scale length $b_*$, ellipticity $\epsilon$, position angle  $PA$ (see Sec.\ \ref{sec:pm}); line-of-sight (los) systemic velocity $V_\te{sys}$,  los velocity  dispersion $\sigma_\te{los}$ estimated considering all the Sculptor's stars within the half-light radius and $1.5^\circ$ (roughly the radial extent of our dataset, see Sec.\ \ref{sec:memb}).
References: 
1) \protect\cite{mc}; 2) \protect\cite{pietr08}; 3) \protect\cite{mateo}; C) Cicu\'endez, private communication;  T) this work. $\dagger$Assuming the distance in the table  the angular to physical scale conversion is  $f_\te{scale}=25.02\pm1.75 \ \pc/\arcmin$. }
\label{tab:obsprop}
\end{table}

The Sculptor dwarf spheroidal galaxy is a satellite of the Milky Way located close to the Galactic South  Pole at Galactic coordinates ($\ell$, $b$)=($287.53^\circ$, $-83.16^\circ$).
%DISTANCE
The distance to Sculptor has been measured with  different methods:  photometry of variable stars (RR Lyrae, e.g.\ \citealt{dRRL}; Miras, e.g.\ \citealt{dMiras}), of horizontal branch stars  (e.g.\ \citealt{dHBS}), using the tip of the red giant branch (e.g.\ \citealt{dTRGB}) and analysing the colour magnitude diagram of the stars in the galaxy (e.g.\ \citealt{dCMD}). 
We decided to take as a reference the value $D=86\pm6 \ \kms$, adopting the distance modulus obtained by \cite{pietr08} using the near-infrared photometry ($J$ and $K$ bands) of a sample of 76 RR Lyrae stars. This value is compatible with the recent estimate by \cite{Garofalo18}.  

There is evidence that Sculptor contains two distinct  distinct stellar  populations: a more concentrated metal rich population with low velocity dispersion ($\approx 6-7 \ \kms$) and a more extended metal poor population with high velocity dispersion ($\approx 10-12 \ \kms$); see e.g.\ \cite{tol04, battagliasculp,w11}. 
As well known,  the existence of two populations can be exploited to constrain the mass distribution of the galaxy (e.g.\ \citealt{battagliasculp,amorisco12,w11,strigari17}).
However, since the purpose of this work is not a detailed study of the internal dynamics of Sculptor, we assume for simplicity the presence of a single stellar component. We discuss the possible implications of this assumption in Sec.\ \ref{sec:dicussion}.

The properties of the light distribution have been derived employing the same state-of-the-art technique used by  \cite{sextans} for the Sextans dwarf spheroidal   applying it on a publicly available VST/ATLAS $g$-band and $r$-band photometric catalogue of stellar sources covering $\approx$ 4 deg$^2$ on the line of sight to Sculptor (Cicu\'endez, private communication).
The best-fit (constant) ellipticity is $\epsilon=0.226\pm0.014$ (axial ratio $q\approxeq 0.77$)  and the best fit position angle is $PA=97.4^\circ\pm1.2^\circ$,  in agreement with the values of \cite{battagliasculp}. 
The best fit to the observed surface number-density profile  is a (projected) Plummer profile
\begin{equation}
N_*\propto\left(1+ \frac{R^2_\te{ell}}{b^{2}_*}\right)^{-2},
\label{eq:obsprofile}
\end{equation} 
where $R_\te{ell}$ is the elliptical radius  corresponding the to major axis of the elliptical iso-density contours and the best fit  scale length is $b_*=11.15\arcmin \pm 0.18\arcmin=279\pm5 \ \te{pc}$. 
The main observed properties of Sculptor are summarised in Tab.\ \ref{tab:obsprop}.

\subsection{Proper motions} \label{sec:pm}

\begin{figure}
\centering
\centerline{\includegraphics[width=1.0\columnwidth]{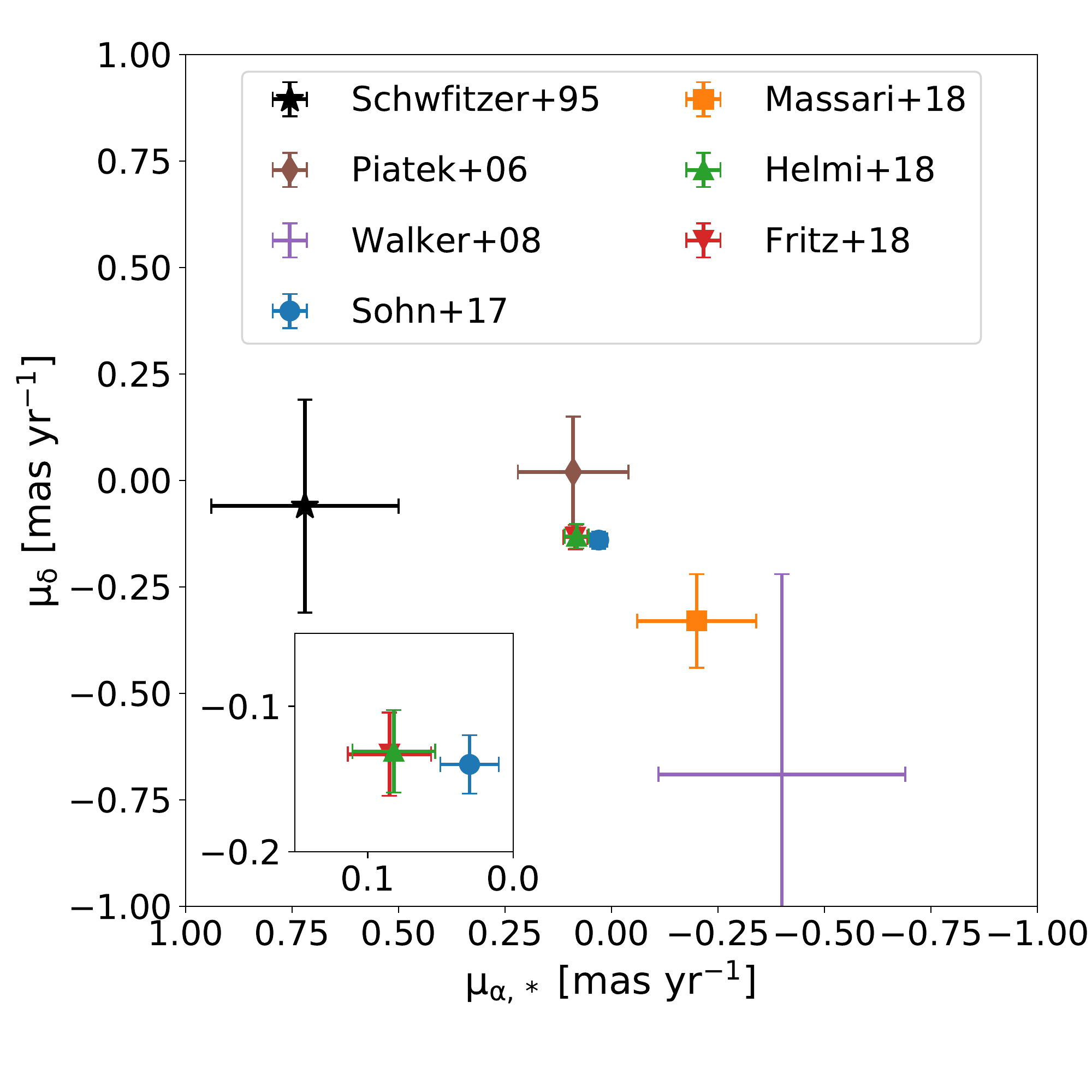}}
\caption[]{
Sculptor's proper motion estimates: \cite{sw95} (black star), \cite{p06} (brown diamond),  \cite{w08} (violet cross), \cite{w11} (pink pentagon),  \cite{s17}  (blue dot),  \cite{ma18} (orange square), \cite{h18} (green up-triangle), \cite{f18} (red down-triangle). The inset shows a zoom-in in the region around the recent \gaia DR2 proper motion estimates.
Note that $\mu_{\alpha,*}=\mu_\alpha \cos \delta$.}
\label{fig:pmots}
\end{figure}

%Old Figure
\begin{comment}
\begin{figure}
\centering
\centerline{\includegraphics[width=1.0\columnwidth]{image/properMotions_recent.pdf}}
\caption[]{Most recent Sculptor's proper motions estimates: \cite{s17}  (blue dot),  \cite{ma18} (red square), \cite{h18} (green up-triangle), \cite{f18} (red down-triangle). Note that $\mu_{\alpha,*}=\mu_\alpha \cos \delta$.}
\label{fig:pmots}
\end{figure}
\end{comment}

There are several direct estimates of the systemic proper motion of Sculptor from astrometric measurements \citep{sw95,p06,s17,ma18,h18,f18} and  indirect estimates by \cite{w08, w11} using the apparent velocity gradient along the direction of the proper motion. 
The currently available estimates of the Sculptor's proper motion are summarised in Fig.\ \ref{fig:pmots}.
While the first measurements were quite uncertain and not all in agreement among each other, in the last couple of years the long time baseline of the HST data and the high-quality astrometric data from \gaia \citep{Gaia,GaiaDR2} have drastically improved the proper motion estimate for Sculptor.  In particular, those  obtained using the $\gaia$ DR2 astrometry (\citealt{h18}; \citealt{f18})  and the one obtained from the HST with a $\approx11$ yr long baseline \citep{s17}  converge toward a likely \vir{definitive} value. 
In the rest of the paper we consider  only  the four most recent estimates: \cite{s17} (S17), \cite{ma18} (M18), \cite{h18} (H18) and \cite{f18} (F18). 
We assumed the errors for the \HP and \citetalias{f18} proper motions (see Fig.\ \ref{fig:pmots} and Tab.\ \ref{ch3:tab:pmres}) as the quadratic sum of  the nominal errors reported in the corresponding papers and the \gaia DR2 proper motions systematic error $\delta \mu _\te{s} \approx0.028\ \te{mas} \ \te{yr}^{-1}$. This last value  has been estimated from Eq.\ 18 in \cite{GaiaDR2} considering a spatial scale of $1^\circ$ roughly corresponding to the extension of the bulk of the \gaia DR2  Sculptor's member analysed in \HP (see Sec.\ \ref{sec:dicussion} for a discussion about the assumed systematic errors). 
Considering the older proper motion estimates it is interesting to compare the two non-astrometric measurements  \citep{w08,w11} with the most recent ones. In these older works the Sculptor's proper motion was estimated assuming that any possible line-of-sight velocity gradient is entirely caused by the relative motion between Sculptor and the Sun (see Sec.\ \ref{sec:binned}). However, if a significant intrinsic rotation is present, the results of this analysis can be heavily biased. Therefore, a significant difference with respect to the astrometric proper motion measurements could be an indication of an intrinsic rotation. We found that the non-astrometric  measurement are consistent within $\approx1 \sigma$ with \MP and within $\approx2 \sigma$ with \SP, \HP and \FP.
Even if the differences are not significant, we note that these results are mainly driven by the large errors reported in \cite{w08,w11}. Hence, a strong claim in favour or against the presence of intrinsic rotation cannot be made from a simple comparison of astrometric and non-astrometric proper motion measurements.

%% file: include/Sculptor/VdispProfile.tex
\subsection{Internal kinematics} \label{sec:vdisp_prof}

\subsubsection{The sample}

We derive the line-of-sight (los) velocity dispersion profile of Sculptor using a state-of-the-art dataset obtained combining the MAGELLAN/MIKE catalogue presented  in \cite{w09data} (1541 stars, \WA hereafter) and the VLT/FLAMES-GIRAFFE catalogue \cite{bs12} (1073 star, \BS hereafter).
The \BS catalogue contains stars  from the original catalogues by \cite{tol04}, \cite{battagliasculp} and \cite{Stark10}.

\subsubsection{Member selection} \label{sec:memb}

\begin{figure}
\centering
\centerline{\includegraphics[width=0.8\columnwidth]{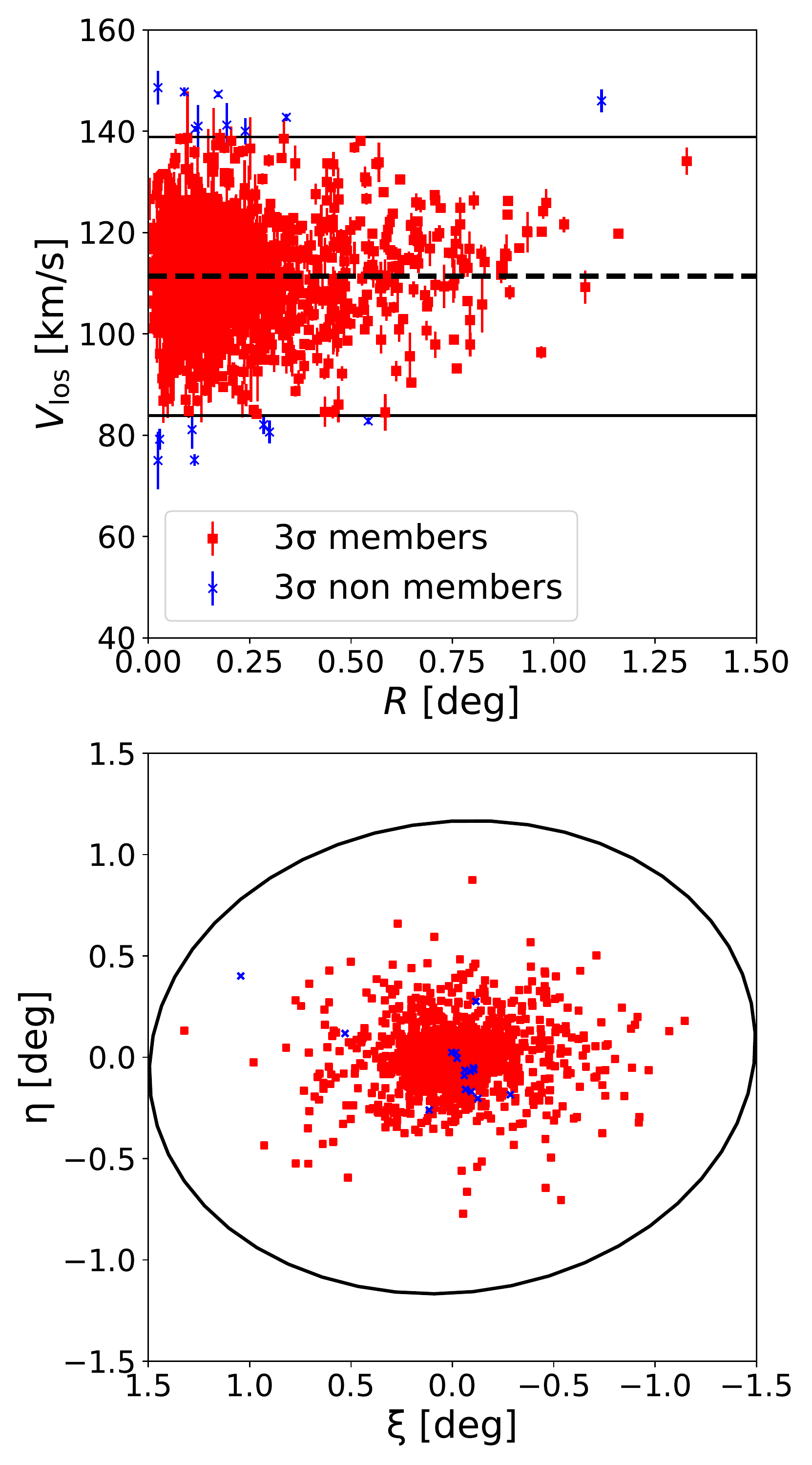}}
\caption[]{Sculptor's member selection. Sculptor's members and  contaminants (likely unresolved binaries) are shown as red squares and blue x-crosses, respectively. Top panel: 
line-of-sight velocity, $\vlos$, versus circular projected radius, $R$; the horizontal lines denote the mean systemic velocity $\vsys$ (dashed) of Sculptor and the $\pm3\sigma$ region (solid).  Bottom panel: position  of the stars with respect to the Sculptor centre ($\xi$ and $\eta$ represent the gnomonic projection of the Galactic sky coordinates); the ellipse is centred on the Sculptor centre: it has a major axis $a=1.5^\circ$, ellipticity $\epsilon=0.226$ and position angle  $PA=97.4^\circ$ (see Sec.\ \ref{sec:obs} and Tab.\ \ref{tab:obsprop}).}
\label{fig:smemb}
\end{figure}

The two catalogues contain both genuine Sculptor's members and Milky Way interlopers.
To start with, for the stars in   \BS, we applied the cleaning of contaminants using the equivalent width of the  Mg I line at $8806.8$ \AA \ as described in \BS.  For the stars in \WA, we removed all the objects with a member probability lower than 0.5 (see \WA for details on the Sculptor's member probability estimate).
Then, we cross-matched the spectroscopic sample with the \gaia DR2 catalogue \citep{GaiaDR2} to complement it with proper motion and parallax estimates.  We removed all the stars that are not compatible with the \HP Sculptor proper motion and with zero parallax  within three times the error. For the parallax, the error corresponds to the uncertainty on the individual parallax measurement; for the proper motion the error is a quadratic sum of the uncertainty in the individual measurement, the error on the systemic proper motion and the systematic error of 0.028 mas/yr \citep[see][see also Sec.\ \ref{sec:pm}] {Lind18}.  
Finally, we removed all the stars with los velocities $V_{\rm los}$  differing at least by $50\ \kms$ from $111\ \kms$  (roughly the Sculptor systemic velocity).
After the selection cuts, we are left with 540 stars from   \BS and 1312 stars from \WA.

In order to identify and analyse duplicate objects, we cross-matched the two catalogues considering a search-window of $1\arcsec$ and we found 276 duplicated stars. 
For each of them we estimate the velocity difference $\Delta V_{\rm los}= V_{\rm los, BS12} - V_{\rm los, W09}$ and the related uncertain $\delta \Delta V_{\rm los}= \sqrt{ \delta V^2_{\rm los, BS12} +  \delta V^2_{\rm los, W09}}$.  We considered that all the stars with  $|\Delta V_{\rm los}/\delta \Delta V_{\rm los}|>3$ are unresolved binary systems and we filtered them out  from the two catalogues.  Among the duplicated stars, only 17 have been removed, resulting in a $6 \%$ fraction of unresolved binaries.
Considering the  los velocity distribution of the 259 remaining stars, we 
found and correct a  velocity offset of $\simeq - 1 \kms$  of the stars in   \BS with respect the stars in \WA. Finally, these stars are combined in a single dataset  estimating the $V_{\rm los}$ as the error-weighted mean of the single velocity measurements.

The joined Sculptor's stars catalogue contains 1559 stars ($\approx 66 \%$ from \WA, $\approx 17 \%$ from  \BS and $\approx 17 \%$ from both  catalogues).  
As a final conservative cleaning criterion, we applied an iterative $3\sigma$ clipping removing 16 stars  with large deviation from the systemic velocity. Fig.\ \ref{fig:smemb} shows that  most of these stars  are located in the central part of Sculptor, hence it is unlikely that they are Milky Way contaminants survived to the selection cuts. Rather, we considered these stars (with single observations in \BS or \WA) as likely unresolved binaries. The fraction of the removed objects ($\approx 1 \%$) is roughly compatible with the binary fraction estimated analysing the stars duplicated among the two catalogues.  
We stress that the $\sigma$ clipping method used to filter binary systems   is capable to select only  short-period binaries (few days; see \citealt{binfrac, spencer18}) for which the typical orbital velocities  are significantly higher than the los velocity errors ( $\langle \delta V_\te{los} \rangle \approx 2 \ \kms$)  and/or than  the dwarf velocity dispersion ($\sigma_\te{los}\approx 9 \kms$, see Fig.\ \ref{fig:compVdisp}). Given the expected distribution of binary periods \citep{Duquennoy91, binfrac},  the short-period systems represent only a tiny fraction of the binaries. Therefore, our estimated binary fraction represents a strong lower limit (\citealt{binfrac} estimated a 0.6 binary fraction for Sculptor). 
However, even if most of the stars in our final sample are in fact part of binary systems, their  expected radial velocities due to the orbital motions are smaller  or compatible with the velocity errors and much smaller than the intrinsic velocity dispersion of the system. Therefore, the kinematic bias introduced by the binary stars on the estimate of the velocity dispersion profile (Sec.\ \ref{sec:binned}) and on the dynamical mass  (Sec.\ \ref{sec:dynfit}) is negligible \citep[see e.g.][]{Hargreaves96,Olszewski96}.

%considering the  $3\sigma$  applied the stars in common between \BS and \WA,  and the dwarf velocity dispersion ($\sigma_\te{los}\approx 9 \kms$, see Fig.\ \ref{fig:compVdisp}) considering the  $3\sigma$ clipping applied to the final joined samples (Fig.\ \ref{fig:smemb}). The values of $\langle \delta V_\te{los} \rangle \approx 2 \ \kms$ and $\sigma_\te{los}\approx 9 \kms$ (see Fig.\ \ref{fig:compVdisp}) set the upper limit of the period of detectable binaries to   $\approx 0.9 \ \te{yr}$ and  $\approx 3.4 \ \te{days}$, respectively   (Eq.\ 17 in \citealt{binfrac}, see also \citealt{spencer18}). 

%Given the expected distribution of binary periods \citep{Duquennoy91, binfrac},  the short-period systems represent only a tiny fraction of the binaries. Therefore, our estimated binary fraction represents a strong lower limit (\citealt{binfrac} estimated a 0.6 binary fraction for Sculptor). 

The final catalogue contains 1543 objects. The systemic velocity of Sculptor, estimated as the mean velocity of its members, is $V_{\rm sys}=111.6\ \pm 0.2\ \kms$; the velocity dispersion is $\sigma_\te{los}=8.9\ \pm  0.2\ \kms$ considering the whole catalogue and $\sigma_\te{los}=8.4\ \pm  0.2\ \kms$ considering only stars  within the half-light radius. The errors on these values have been estimated using the bootstrap technique \citep{Feigelson12}.

\subsubsection{Velocity dispersion profile} \label{sec:binned}

The relative motion of Sculptor with respect to the Sun causes an artificial los velocity gradient  along the proper motion direction \citep[see e.g.\ ][]{w08}. Therefore,  we corrected the measured los velocity for this perspective effect through the methodology described in the Appendix A of \cite{w08} using the systemic proper motion estimated in \HP and the systemic los velocity reported in Tab.\ \ref{tab:obsprop}. The errors on the corrected $V_\te{los}$ have been estimated with a Monte Carlo simulation considering the observational uncertainties on $V_\te{los}$, the errors on $V_\te{sys}$ and distance (Tab.\ \ref{tab:obsprop}), and the error on  the proper motions (Tab.\ \ref{ch3:tab:pmres}).

\begin{figure}
\centering
\centerline{\includegraphics[width=1.1\columnwidth]{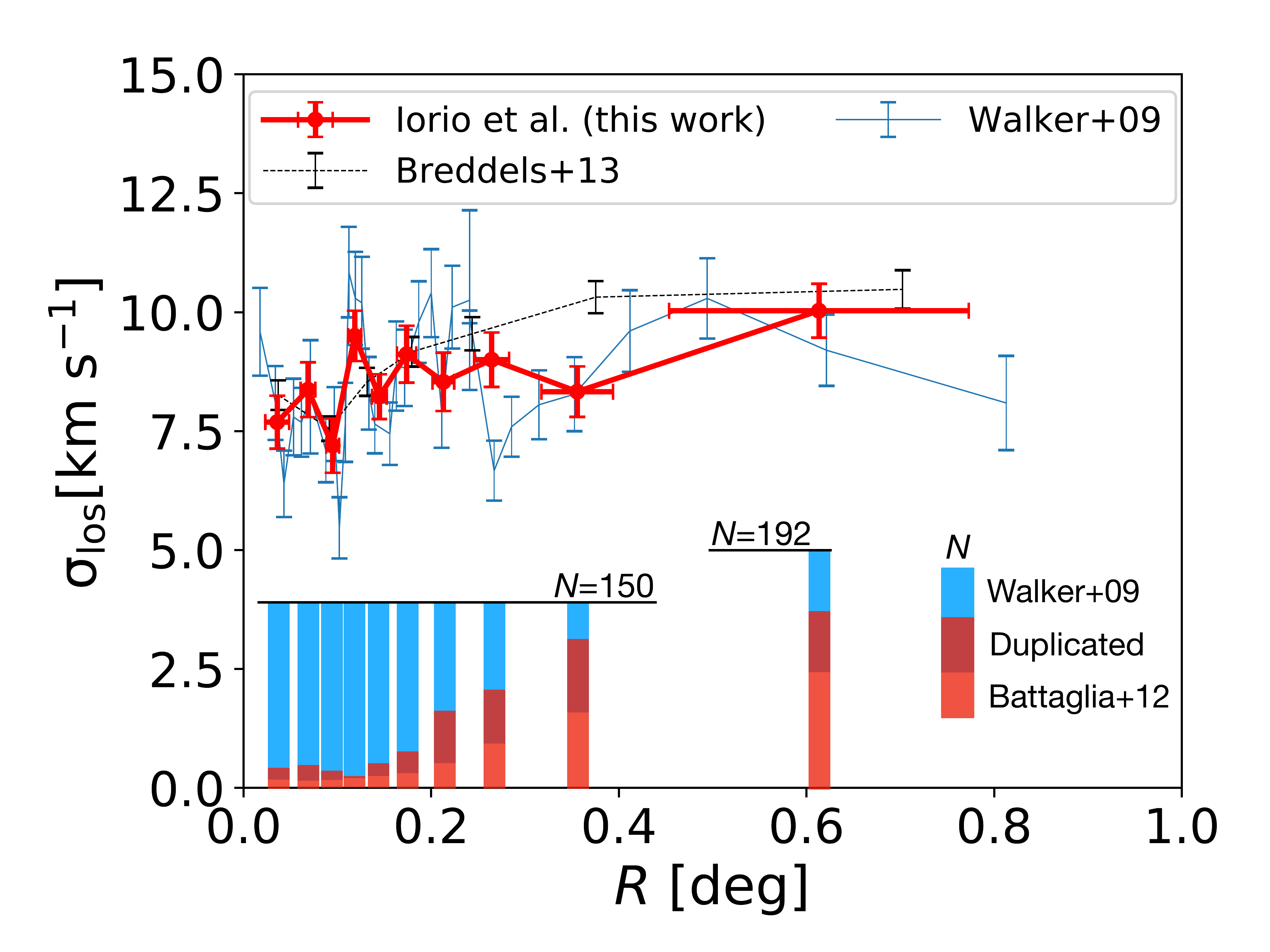}}
\caption[]{Velocity dispersion profile of Sculptor: this work (red points and line), \cite{w09} (blue line), \cite{breddelsdsph} (black dashed line). The histograms in the bottom part of the plot indicate the fraction of stars in each bin belonging to the original samples of \WA (blue bars) and of \BS (red bars). The dark-red areas, coming from the intersection of the blue and red bars, represent the fraction of common stars between the two samples (see Sec.\ \ref{sec:memb}). All bins have 150 stars, but the outermost bin that has 192 stars. See the text for the detailed description of our binning technique.}
\label{fig:compVdisp}
\end{figure}

The radial profile of los velocity dispersion $\sigma_\te{los}(R)$,  has been obtained grouping Sculptor's members in (circular) radial bins. In order to have the best compromise between good statistics and spatial resolution (bin width), we decided to select the bin edges so that each bin contains 150 stars. Since the outermost bin remains with  41 stars only, we merged it with the adjacent bin, obtaining a wider outermost bin with 192 stars. 
For each of the final 10 bins, we estimated  the   mean los velocity $\overline{V}_{\te{los}}$  and the velocity dispersion $\sigma_\te{los}$  fitting a Gaussian model  and  taking into account the $V_\te{los}$ uncertainties through an extreme-deconvolution technique \citep{bovyxdd}. Similarly, we estimated the representative radius of each bin taking the mean of the circular radius of the stars. The errors on these values have been estimated using the bootstrap technique \citep{Feigelson12}.
The resultant $\sigma_\te{los}(R)$ profile is shown and compared with the profiles from  \cite{w09} and \cite{breddelsdsph} in Fig.\ \ref{fig:compVdisp}. 
The  two profiles  are compatible with our profile within the half-light radius.  Beyond this radius the $\sigma_\te{los}$  estimated in this work are systematically  lower with respect to the values of \cite{breddelsdsph}, but  compatible with the \cite{w09} estimates. 
This discrepancy is not due to the predominance of stars from \WA in our joined catalogue (see Sec.\ \ref{sec:memb}), because the two outermost bins are dominated by stars taken from \BS (see bottom bars in Fig.\ \ref{fig:compVdisp}). The most likely explanation is that the velocity dispersion profile reported in \cite{breddelsdsph} is slightly inflated by residual Milky Way interlopers. 

The  assumptions we made to obtain the final binned profile are discussed below.
\begin{itemize}
\item {$N_\te{star}$} per bin. We repeated the analysis described above changing the number of stars per bin from 100 to 350 in steps of 50. The increase of the number of stars per bin reduces the errors in $\sigma_\te{los}$ but it decreases the spatial resolution. The final profiles are all compatible, but the chosen value of 150 represents the best compromise between the two kinds of uncertainties.
\item {Velocity gradient.}  The correction for the perspective velocity gradient due to the relative motion between the Sun and Sculptor has been applied using the \HP proper motion. We repeated the correction procedure considering also the \MP and \SP proper motions. The differences on the final velocity dispersion profiles are negligible ($<1 \%$). Even if some residual shallow velocity gradient should still be present, e.g. due to an intrinsic rotational motion (e.g. \citealt[][see also Sec.\ \ref{sec:pm}]{battagliasculp,zhu16}), any possible bias in the velocity dispersion should be important only in the outermost bins where the ratio between the velocity dispersion and the rotation velocity is larger. Considering as a test a rotation velocity of $7 \ \kms$ and a velocity dispersion of $10 	\ \kms$ and a sample of 150 stars uniformly distributed along the azimuthal angle, we found that the increase of the velocity dispersion is expected to be lower than $10\%$. This value is \vir{safely}  within the error bars of the outermost points of the estimated velocity dispersion profile.  

\item {Circular vs elliptical bins.}  
We estimated new velocity dispersion profiles as a function of  the elliptical radius $R_\te{ell}=\sqrt{x^2+y^2/q^2}$ and of the circularised radius $R_\te{circ}=\sqrt{x^2q+y^2/q}$, where  $q$ is the observed Sculptor's axial ratio (see Sec.\ \ref{sec:obs}) and $(x,y)$ are the  Cartesian coordinates in the reference aligned with the Sculptor's principal axes. The $\sigma_\te{los}(R_\te{ell})$ and $\sigma_\te{los}(R_\te{circ})$ profiles are practically coincident and both the normalisation and the radial trend are compatible with the los velocity dispersion profile shown in Fig.\ \ref{fig:compVdisp}.

\end{itemize}

We stress that the aim of this paper is not the detailed study of the velocity dispersion profile  in Sculptor, but rather we want to have a self-consistent way to infer the velocity dispersion profile both in observations and in simulations. The performed analysis and the comparison of our profile with results in the literature (Fig.\ \ref{fig:compVdisp}) ensures that our  \vir{simple} technique and our assumptions do not cause severe biases in the final estimate of  $\sigma_\te{los}$.

%% file: include/Orbital/Orbital.tex
\section{The orbit of Sculptor} \label{sec:orbital}

\subsection{Milky Way model} \label{sec:orbital:mw}

\begin{table*}
\centering
\begin{tabular}{ccc|cc|cc|}
\cline{4-7}
\multicolumn{1}{l}{}                             & \multicolumn{1}{l}{}                                                                      & \multicolumn{1}{l|}{}                                                                     & \multicolumn{4}{c|}{Milky Way model}                                                                                                                                                                                                                                                              \\ \hline
\multicolumn{1}{c|}{}                            & \multicolumn{2}{c|}{Proper motions}                                                                                                                                                   & \multicolumn{2}{c|}{\BP}                                                                                                                         & \multicolumn{2}{c|}{\JP}                                                                                                                       \\ \hline
\multicolumn{1}{c|}{Model}                       & \begin{tabular}[c]{@{}c@{}}$\mu_{\alpha,*}$\\ ($\te{mas} \ \te{yr}^{-1}$)\\ (1a)\end{tabular} & \begin{tabular}[c]{@{}c@{}}$\mu_\delta$\\ ($\te{mas} \ \te{yr}^{-1}$)\\ (1b)\end{tabular} & \begin{tabular}[c]{@{}c@{}}$\rapo$\\ {[}kpc{]}\\ (2a)\end{tabular}           & \begin{tabular}[c]{@{}c@{}}$\rperi$\\ {[}kpc{]}\\ (2b)\end{tabular}     & \begin{tabular}[c]{@{}c@{}}$\rapo$\\ {[}kpc{]}\\ (3a)\end{tabular}        & \begin{tabular}[c]{@{}c@{}}$\rperi$\\ {[}kpc{]}\\ (3b)\end{tabular}      \\ \hline
\rowcolor[HTML]{CCCCCC}  
\multicolumn{1}{c|}{\cellcolor[HTML]{CCCCCC}\HP} & $0.082\pm0.028^\dagger$                                                                           & $-0.131\pm0.028^\dagger$                                                                          &       \begin{tabular}[c]{@{}c@{}}$145^{+17}_{-23}$\\ $[99,246]$\end{tabular}                                                                      &                \begin{tabular}[c]{@{}c@{}}$65^{+6}_{-7}$\\ $[45,83]$\end{tabular}                                                       & \begin{tabular}[c]{@{}c@{}}$105^{+8}_{-7}$\\ $[84,132]$\end{tabular}   & \begin{tabular}[c]{@{}c@{}}$54^{+6}_{-4}$\\ $[36, 74]$\end{tabular}   \\
\multicolumn{1}{c|}{\FP}                         & $0.085\pm0.029^\dagger$                                                                           & $-0.133\pm0.029^\dagger$                                                                          &  \begin{tabular}[c]{@{}c@{}}$143^{+17}_{-22}$\\ $[99,244]$\end{tabular}                                                                         &                    \begin{tabular}[c]{@{}c@{}}$64^{+6}_{-6}$\\ $[45,83]$\end{tabular}                                                   &  \begin{tabular}[c]{@{}c@{}}$105^{+7}_{-8}$\\ $[83,130]$\end{tabular}   & \begin{tabular}[c]{@{}c@{}}$54^{+6}_{-7}$\\ $[36, 74]$\end{tabular}   \\
\rowcolor[HTML]{CCCCCC}  
\multicolumn{1}{c|}{\cellcolor[HTML]{CCCCCC}\MP}                        & $-0.200\pm0.140$                                                                          & $-0.330\pm0.110$                                                                          & \begin{tabular}[c]{@{}c@{}}$361^{+425}_{-273}$\\ $[81,2428]$\end{tabular} & \begin{tabular}[c]{@{}c@{}}$81^{+7}_{-9}$\\ $[21, 101]$\end{tabular} & \begin{tabular}[c]{@{}c@{}}$141^{+49}_{-55}$\\ $[73,811]$\end{tabular} & \begin{tabular}[c]{@{}c@{}}$78^{+11}_{-11}$\\ $[18,101]$\end{tabular} \\
\multicolumn{1}{c|}{\SP} & $0.030\pm0.020$                                                                           & $-0.140\pm0.020$                                                                          & \begin{tabular}[c]{@{}c@{}}$152^{+17}_{-21}$\\ $[111,266]$\end{tabular}   & \begin{tabular}[c]{@{}c@{}}$67^{+5}_{-11}$\\ $[54, 85]$\end{tabular} & \begin{tabular}[c]{@{}c@{}}$108^{+8}_{-8}$\\ $[87,135]$\end{tabular}   & \begin{tabular}[c]{@{}c@{}}$60^{+5}_{-6}$\\ $[45, 76]$\end{tabular}   \\
\hline
\end{tabular}
\caption{Orbital parameters estimated for different proper motions  of Sculptor: \HP, \FP, \MP and \SP (Cols. 1a/1b). For each of them, we integrated $10^5$ orbits varying all the Sculptor (Tab.\ \ref{tab:obsprop}) and \MW parameters accordingly with their errors (see the text). The columns report the statistics of the posterior distributions obtained for the radius of the apocentre (Cols. 2a/3a) and of the pericentre (Cols. 2b/3b) assuming the Milky Way models \BP and \JP. The  numbers indicate the
values obtained considering the fiducial value for the  Sculptor and Milky Way parameters, the superscript and the subscript represent the interval that contains the $68\%$ ($1\sigma$) of the distribution. The   values inside the square brackets indicate the interval containing the $99.7\%$ ($3\sigma$)  of the distribution. $\dagger$The errors are the quadratic sum of the nominal errors reported in \HP and \FP and the $0.028\ \te{mas} \ \te{yr}^{-1}$ \gaia systematic error \citep{Lind18}. }
\label{ch3:tab:pmres}
\end{table*}

Despite a large number of studies focusing on the Milky Way dynamics, the detailed gravitational potential of the Galaxy is still far from being nailed down.
Although  there is convergence in the mass estimate in the inner part of the DM halo 
\citep[$r\lesssim50\ $ kpc, e.g.][]{mww99,mwxue08,mwd12,mwg14},  results of different works predict a relative wide  range of values for the Galactic virial mass\footnote{The virial mass $M_{200}$ is the mass within the virial radius $r_{200}$, such that  
the mean halo density within $r_{200}$ is 200 times the critical density of the Universe.} $M_{200}$ \citep[][and reference therein]{wang15}. 
%Moreover,  even the estimates in the same paper can vary a lot depending on the working assumptions (e.g.\  concentration parameter as in \citealt{mwb05} or anisotropy profile of the tracers as in \citealt{mwg10}).
We can roughly divide the different results in three  groups\footnote{Different works can have different definition for the MW mass. When we refer to literature values   we consider the conversion to   $M_{200}$ reported in \cite{wang15}}: works predicting a \vir{heavy} 
Milky Way with $M_\te{200}\gtrsim2\times10^{12} \ \msun $ \citep[e.g.][]{li08}, works predicting a \vir{light} Milky Way with $M_\te{200}\lesssim1\times10^{12} \ \msun$   \citep[e.g.][]{mwb05,mwg14} and other works in between with $M_\te{200}\approx 1.2-1.8 \times10^{12} \ \msun$ \citep[e.g.][see also recent estimates by \citealt{Cortigiani18, MWMonari18, mwp18}]{mwp14}. 

In this work we considered two \vir{extreme} analytical Milky Way dynamical models: the \vir{heavy} model by \citet[][\JP]{j95}, and the \vir{light} model by \citet[][\BP]{b14}. We decided to investigate the orbital properties of Sculptor both in the \JP and \BP potentials.
The \JP model consists in a Miyamoto-Nagai disk \citep{mnd} with mass  $M=10^{11} \ \msun$, a spherical Hernquist  bulge \citep{hrn} with mass $M=3.40\times10^{10} \ \msun$ and a spherical logarithmic dark halo. 
The \BP model includes a Miyamoto-Nagai disk with mass $M=6.80\times10^{10} \ \msun$, a spherical bulge with a truncated power-law density profile with mass $M=0.45\times10^{10} \ \msun$ and a  spherical Navarro-Frenk-White (NFW; \citealt{nfw}) dark halo.
The mass profile in \JP rises from $\simeq 0.5\times10^{12} \ \msun$ at $50 \ \kpc$
to  $\simeq 2.3\times10^{12} \ \msun$ at $300 \ \kpc$. 
In the same radial range the mass in \BP increases from  $\simeq 0.3\times10^{12} \ \msun$ to  $\simeq 1.0 \times 10^{12} \ \msun$.

We set a left-handed Cartesian reference frame
$(x_\te{g},y_\te{g},z_\te{g})$ centred in the Galactic
centre and such that the Galactic disc lies in the $x_\te{g}$-$y_\te{g}$ plane, the Sun lies on the positive
$x$-axis   and the Sun rotation velocity  is
$\dot{y}_\te{g}>0$.  In this frame of reference,  the Sun is at a distance of $R_\odot=8.0\pm0.5$ kpc \citep{rsunbovy} from the Galactic centre and  the rotational velocity of the local standard of rest (lsr) is  $V_\te{lsr}=218.0\pm6.0 \ \kms$  \citep{rsunbovy}.
 Finally, we used the estimate of the solar proper motion with respect to the lsr reported in \cite{rsunsch}:  ($U_\odot,\ V_\odot,\ W_\odot$)=($-11.1\pm1.3,\ 12.2\pm2.1,\ 7.25\pm0.6$) $\kms$. Therefore, the motion of the Sun in the Galactic frame of reference is $\vec{\upsilon}_\odot=(U_\odot, \ V_\odot+V_\te{lsr}, \ W_\odot)$. %Considering the assumed Galactic frame of reference,  a negative $U_\odot$ means that the motion of the Sun is towards the Galactic centre, a positive $V_\odot$ is towards the direction of the Galactic rotation and a positive $W_\odot$ is towards the North Galactic Cap. 
The estimated  solar motion ($\vec{\upsilon}_\odot$) is used in Sec.\ \ref{sec:orbinv} to obtain the 
velocity of Sculptor in the Galactocentric frame of reference   from the observed kinematics (proper motions and los velocity).

\subsection{Orbital parameters} \label{sec:orbinv}

We used the leapfrog orbit integrator in the \texttt{Python} module \texttt{galpy} \citep{galpy} to investigate the orbits of Sculptor. Combining the  the proper motion estimates (\HP, \FP, \MP and \SP, see Sec.\ \ref{sec:pm}) and the two Milky Way potentials (\JP and \BP; see Sec.\ \ref{sec:orbital:mw}) we consider 8 families of orbits. For each family, we performed $10^5$ orbit integrations drawing the values of the proper motions, the Sculptor systemic velocity, the Sculptor distance, the Sun position and motion in the Galactic frame of reference from Gaussian distributions centred on the measured values and with dispersion given by the corresponding errors. 
In the case of \HP and \FP, we took into account the reported proper motion correlation extracting the values of $\mu_{\alpha,*}$ and $\mu_\delta$ from a bidimensional normal distribution. 
Each orbit has been integrated backward from the present-day position of Sculptor for $8$ Gyr.
The chosen integration time represents a  good compromise between a time long enough to explore the effect of tides and a time window in which the mass evolution of the Milky Way can be considered almost negligible. In particular, we notice that the  last significant Milky Way merger  likely happened  8-11 Gyr ago \citep{BelokurovSa,MyGC, Helmi18,Kraken}.

The final apocentre and pericentre distributions for each of the 8 investigated families of orbits are summarised in Tab.\ \ref{ch3:tab:pmres}.  The \HP, \FP and \SP proper motion estimates give very similar results, largely compatible within the error. As expected the orbits in the \vir{light} Milky Way potential (\BP) have both larger apocentre and pericentre with respect to the case of the \vir{heavy} Milky Way model (\JP).
The results obtained for \MP using the fiducial proper motion values are slightly offset towards more external orbits, however given the larger proper motion errors, the overall results are still compatible with the other estimates.  
It is worth noting that the pericentres found assuming  the \HP, \FP and \SP proper motions are consistent  with the first estimate of the Sculptor's pericentre, $\rperi=68 \ \kpc$, obtained by \cite{p06}.

\begin{table}
\centering
    \tabcolsep 4pt
    %\small
\begin{tabular}{c|ccccc}
\hline
     & \begin{tabular}[c]{@{}c@{}c@{}}$R_\odot$\\ \\ {[}kpc{]} \\ (1)\end{tabular} & \begin{tabular}[c]{@{}c@{}c@{}}$\vec{v}_\odot$\\ $(V_x,V_y,V_z)$\\ {[}$\kms${]}\\ (2) \end{tabular} & \begin{tabular}[c]{@{}c@{}c@{}}$V_\te{sys}$\\ \\ {[}$\kms${]} \\ (3) \end{tabular} & \begin{tabular}[c]{@{}c@{}c@{}c@{}}$D$\\ \\ {[}kpc{]} \\ (4) \end{tabular} & \begin{tabular}[c]{@{}c@{}c@{}}$\vec{\mu}$\\ $(\mu_{\alpha,*}, \mu_\delta)$\\ {[}$\te{mas} \ \te{yr}^{-1}${]} \\ (5) \end{tabular} \\ \hline
FH18 & $8.0$                                                            & $(-11.1, 230.2, 7.3)$                                                                    & $111.6$                                                                & $86.0$                                                     & $(0.085,  -0.133)$                                                                                                 \\
EH18 & $8.2$                                                            & $(-10.9, 213.3, 7.9)$                                                                    & $111.8$                                                                & $82.3$                                                     & $(0.125, -0.164)$                                                                                                  \\
%E\MP & $8.2$                                                            & $(-09.8, 231.3, 8.1)$                                                                     & $110.7$                                                                & $92.3$                                                     & $(0.168, -0.360)$                                                                                                  \\ \hline
\hline
\end{tabular}
\caption{Parameters that give rise to the two orbits used in the $N$-body simulations. (1) distance of the Sun from the Galactic centre; (2) proper motion of the Sun in the Galactic frame of reference (see text); (3) Sculptor's los systemic velocity; (4) Sculptor's distance from the Sun; (5) Sculptor's proper motion.   }
\label{tab:simdata}
\end{table}

Given the similarity of the orbital properties, we decided, without loss of generality, to consider the distribution of orbital parameters obtained with the \HP proper motions as our reference to set up of $N$-body simulations. 
Moreover, since we are interested in the influence of  \MW tidal field on the observed properties of Sculptor, we make the conservative choice to consider in the simulations the \JP \MW potential model. If no strong tidal effects are found using this potential, it is likely that we can safely extend this result to any other realistic axisymmetric \MW potential model.
\begin{figure}
\centering
\centerline{\includegraphics[width=1.2\columnwidth]{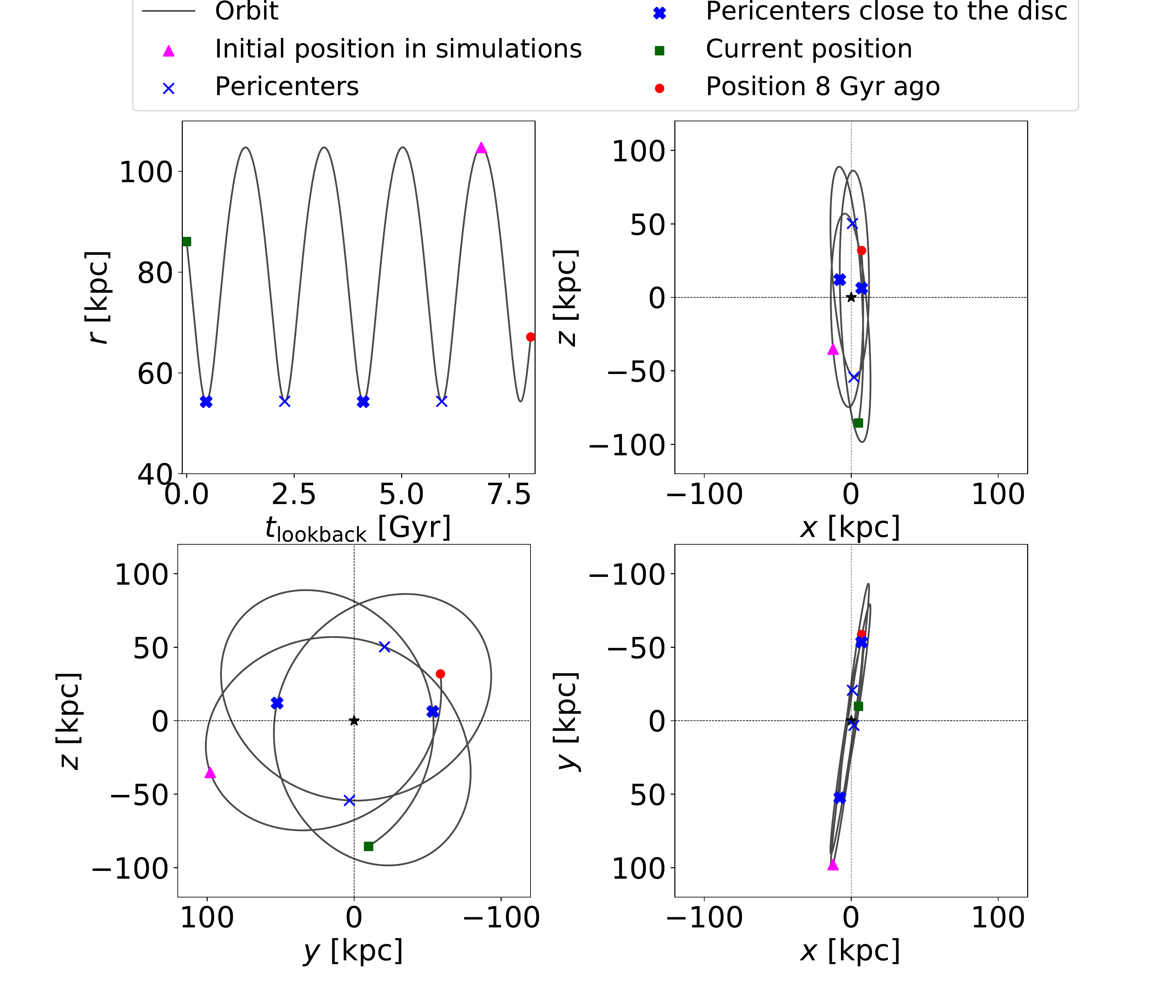}}
\caption[]{FH18 orbit of Sculptor (see Tab.\ \ref{tab:simdata}) integrated backward for 8 Gyr from the present position (green square).  The $(x,y,z)$ Galactocentric frame of reference is defined in Sec.\ \ref{sec:orbital:mw} and $r=\sqrt{x^2+y^2+z^2}$ is the Galactocentric radius.
The lookback time $t_\te{lookback}$ indicates the time in the past considering the backward orbit integration.
The red circle is the position at the end of the orbit integration. The magenta triangles indicate the starting location of Sculptor in the $N$-body simulation (see Sec.\ \ref{sec:icord}). The blue crosses indicate the positions of the pericentres,  the thicker crosses highlight the pericentres located close to the Galactic plane ($z<20\ \kpc$). The black stars indicate the Galactic centre. }
\label{fig:orb_fiducial}
\end{figure}

\begin{figure}
\centering
\centerline{\includegraphics[width=1.2\columnwidth]{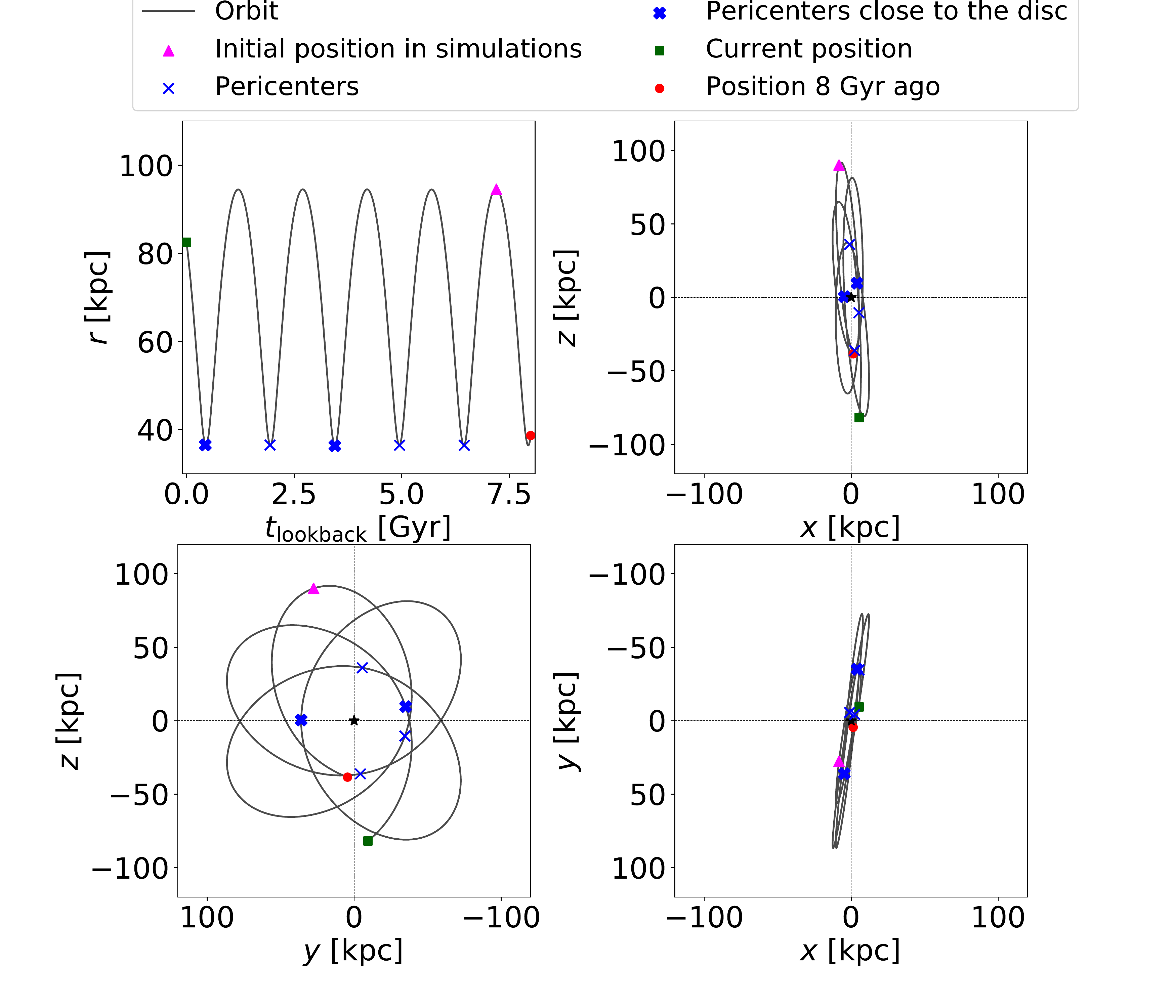}}
\caption[]{Same as Fig.\ \ref{fig:orb_fiducial} bur for the EH18 orbit (see Tab.\ \ref{tab:simdata}).}
\label{fig:orb_extreme}
\end{figure}

In the  $N$-body simulations (Sec.\ \ref{sec:Nbodysim}) we consider two different orbits. 
The first, labelled FH18, has been obtained using the fiducial values of all the Sculptor  and Milky Way parameters reported in Tab.\ \ref{tab:obsprop}, Tab.\ \ref{ch3:tab:pmres} and Sec.\ \ref{sec:orbital:mw}.
The second, labelled EH18, is the orbit that, among all the $10^5$ realisations, has the pericentric radius closest to the left $3\sigma$ extreme of the pericentre distribution.  
The FH18 orbit has  pericentre   $\rperi=54 \ \kpc$,   apocentre  $\rapo=105 \ \kpc$, eccentricity $e=0.32$, and radial orbital period $T=1.6\ \te{Gyr}$.
In the EH18 case, the orbit has  pericentre  $\rperi=36\ \kpc$,   apocentre  $\rapo=94\ \kpc$, eccentricity $e=0.44$, and radial orbital period  $T=1.3\ \te{Gyr}$.
The projections of the FH18 and EH18 orbits in the Galactic planes are shown in Fig.\ \ref{fig:orb_fiducial} and Fig.\ \ref{fig:orb_extreme}, respectively. 
In both the FH18 and EH18 orbits Sculptor is currently heading towards its next apocentre after the last  pericentric passage ($\approx 0.5\ \te{Gyr}$ ago). 
The parameters that give rise to the orbits FH18, EH18 and EM18 are summarised in Tab.\ \ref{tab:simdata}.

%% file: include/Dynamics/DynamicalModel.tex
\section{Dynamical models of Sculptor} \label{sec:dynmod}

In order to set up the initial conditions of the $N$-body simulations (Sec.\ \ref{sec:Nbodysim}) we have to define a density model for both the DM halo and the stellar component of Sculptor. As a startig point, in the present section we construct spherically symmetric equilibrium dynamical models that represent Sculptor as it is observed today.

\subsection{Density distributions}

\subsubsection{Stars} \label{sec:dstar}
Based on the observed light distribution of Sculptor (see Sec.\ \ref{sec:obs}), we used a 3D Plummer profile to model the density law for the stellar component,
\begin{equation}
\rho_*(r) = \rho_0  \left(1+ \frac{r^2}{b^2_\te{*}}\right)^{-2.5}.
\label{eq:d3dplummer}
\end{equation}
In Eq. \ref{eq:d3dplummer} the angular length of the Plummer core radius, $b_*=11.15\arcmin$, is set from the fit to the observed light profile of Sculptor (Sec.\ \ref{sec:obs}) and the physical length depends on the assumed distance that can vary in different orbits (see Tab.\ \ref{tab:simdata}). Thus  $b_*=0.279\ \kpc$ for the FH18 orbit ($D=86.0$ kpc) and $b_*=0.267\ \kpc$ for the EH18 orbit ($D=82.3$ kpc).
The normalisation factor  is 
\begin{equation}
\rho_0=M_* \frac{3}{4\pi b^3_*} \left( \frac{b^2_*+R^2_*}{R^2_*} \right),
\label{eq:normfact}
\end{equation} 
where $R_*=1.5^\circ\simeq8 b_*$ ($\approx 2.2 \ \kpc$ and $\approx 2.1 \ \kpc$  for  FH18 and  EH18 orbits respectively) is a  reference (projected) radius that is  equal to the length of the  major axis of the ellipse containing all the stars in our sample (see Fig.\ \ref{fig:smemb}). 
Given the Plummer surface density profile 
\begin{equation}
\Sigma_*(R) = \rho_0 \int^{\infty}_R \te{d}r  \left(1+ \frac{r^2}{b^2_\te{*}}\right)^{-2.5} \frac{r}{\sqrt{r^2-R^2}}=\Sigma_0 \left( 1 + \frac{R^2}{b^2_*} \right)^{-2},
\label{eq:d2dplummer}
\end{equation}
the value of the parameter $\rho_0$ introduced in Eq.\ \ref{eq:normfact} is fixed by the condition
$2\pi\int_0^{R_*}\Sigma_*(R)R\te{d} R=M_*$, where
$M_*=\Upsilon_* L_*=4.6\times 10^6 \ \msun$ and $L_*=2.3\times10^6$ is the total luminosity of Sculptor (Tab.\ \ref{tab:obsprop}) and  $\Upsilon_*=2$ is the assumed mass-to-light ratio \citep{battagliasculp, woo08}.
The choice of $R_*$  is relatively arbitrary, but  this parameter does not have significant impact  in the  results of our dynamical model fitting (see Sec.\ \ref{sec:dynfit}), given that for $R_* \gg b_*$ the central density (Eq.\ \ref{eq:normfact}) tends to the constant value $ 3 M_* \left( 4\pi b^3_* \right)^{-1}$.

\subsubsection{Dark Matter} \label{sec:dDM}
Pure  $N$-body cosmological simulations in the $\Lambda$ cold dark matter framework produce DM haloes with a double power-law density structure and an inner cusp, the so-called NFW profile \citep{nfw}. However, a number 
of baryonic processes are able to \vir{carve} a core in the original DM cusp \citep[e.g.][]{Pontzen12,nipoticore,readcore}.
Different studies point out that cored DM density profiles provide better fits to kinematic data of dwarf galaxies. This is a strong observational evidence  in dwarf irregulars (e.g.\ \citealt{readiorio17}). It is less clear in dwarf spheroidals (see e.g.\  \citealt{breddelsdsph,zhu16}), though in some cases cuspy haloes are strongly disfavoured by the data (see e.g.\  \citealt{pascale18}, but see also \citealt{readdraco}).  

\begin{comment}
We note that the formation of cores of the size of the half-light radius in "classical" Milky Way dwarf spheroidals is compatible with estimates of the energy input by explosions of supernovae type II based on the observed star formation history of these galaxies \citep{Bermejo-Climent18}.  
\end{comment}

In order to gain the flexibility to include a core while maintaining the general properties of the NFW profile,  we define the density profile of Sculptor's DM halo as
\begin{equation}
\rho_\te{DM}(r)= \rho_\te{crit} \delta_c \left( \frac{r^2_\te{c} + r^2}{r^2_\te{s}}  \right)^{-0.5} \left(  1 +  \frac{r}{r_\te{s}}  	\right)^{-2}.
\label{eq:dNFWc}
\end{equation}
Eq.\ \ref{eq:dNFWc} is a modified version of the NFW density profile with an inner core length $r_\te{c}$ (it gives the NFW law for $r_\te{c}=0$), while the other terms are the usual NFW parameters: $r_\te{s}$ is the radial scale length, 
\begin{equation}
\delta_c=\frac{200}{3} \frac{c^3}{\ln(1+c) -c/(1+c)}
\end{equation} is the concentration factor  that depends on the concentration parameter $c=r_\te{200} r^{-1}_s$. For $r \gg r_\te{c}$, Eq.\ \ref{eq:dNFWc} has the same properties as an NFW density profile. 

In particular, since in our application $r_\te{c}\approx0.05r_\te{s}$ (see Sec.\ \ref{sec:dyn:results})   and $c>10$ (see Tab.\ \ref{tab:result}),  the virial mass, $M_\te{200}=M_\te{DM}(<r_\te{200})$, derived from Eq.\ \ref{eq:dNFWc} is nearly equal to the case of a classical NFW halo with the same $r_\te{s}$ and $c$.
We reduce the number of model parameters using  the semi-analytical concentration-mass relation derived by \citealt{diemer19}. Specifically, we computed the value of the concentration parameter for given virial mass with 
the Python module \texttt{Colossus} \citep{colossus} assuming cosmological model \vir{planck15} \citep{planckpar}, concentration model \vir{diemer19} and redshift $z=0$. 
Therefore, each DM model is fully described with only two parameters: the core radius, $r_\te{c}$, and the virial mass, $M_\te{200}$. 
The concentrations parameter of the dwarf simulated in  this work   ranges approximately between 12 and 15 (see Tab.\ \ref{tab:result}).
Considering the typical virial mass of  dwarf galaxies ($M_{200}\sim10^9-10^{10}\msun$,  see Sec.\ \ref{sec:dyn:results}), the virial radius $r_\te{200}$ can be as large as several tens of kpc. 
Due to the tidal field of the Milky Way (see Sec.\ \ref{sec:nrel}), we do not expect the halo of Sculptor to extend out to $r_{200}$, nonetheless it is useful to consider such kind of theoretical model to determine the expected dark matter distribution within the truncation radius.

%\begin{equation}
%\log c_{200} = -0.00603  \left( \log  \frac{M_\te{200}}{\te{M}_\odot} \right)^2 +0.02967 \log % \frac{M_\te{200}}{\te{M}_\odot} + 1.39955.
%\label{eq:cMvir}
%\end{equation} 
%The relation in Eq.\ \ref{eq:cMvir} is a polynomial fit to the  semianalytical concentration-mass relation derived by \citealt{diemer19}. In particular,  assuming the \cite{planck} cosmological model ($h=0.67$)
%The relation in Eq.\ \ref{eq:cMvir} has been derived for the local Universe (redshift $z=0$) using the  cosmological parameter from \cite{planckpar} (e.g.\ dimensionless Hubble parameter $h=0.67$) and the concentration models from\footnote{The $c_{200}-M_\te{200}$ relation in Eq.\ \ref{eq:cMvir} is  a polynomial fit to the true relation obtained  with the Python module \texttt{Colossus} \citep{colossus} using the cosmological model \vir{planck15}, the concentration model 
%\vir{diemer18} and assuming redshift $z=0$. Eq.\ \ref{eq:cMvir} is  valid (errors lower than 1\%) in the mass interval $(10^7,10^{12}) \ \msun$.} \citealt{diemer18}.

\subsection{Fitting the line-of-sight velocity distribution} \label{sec:dynfit}
We obtained the DM halo parameters for Sculptor ($r_\te{c}$, $M_\te{200}$) fitting the corrected (see Sec.\ \ref{sec:binned}) $V_\te{los}$ distribution  of Sculptor's members (Sec.\ \ref{sec:memb}) through a Jeans analysis.
Assuming spherical symmetry and isotropic velocity dispersion, the velocity dispersion $\sigma_\te{los}$ along the line of sight and the total matter distribution $M_\te{tot}(r)$ are related by \citep{jeans}
\begin{equation}
\sigma^{2}_\te{los}(R)=2G \Sigma^{-1}_{*}(R) \int^{\infty}_{R} \frac{\te{d}r}{r} \rho_{*}(r) M_\te{tot}(r)   \sqrt{1 -\frac{R^{2}}{r^{2}}},
\label{eq:jeans}
\end{equation}
where $M_\te{tot}(r)$, and thus $\sigma_\te{los}(R)$, depend on $r_\te{c}$ and $M_\te{200}$.
In Eq.\ \ref{eq:jeans},  $\Sigma_{*}$ is the observed surface density profile (Eq.\ \ref{eq:obsprofile}) and $\rho_{*}$ the  3D stellar density distribution, which in our case is represented by the Plummer profile (Eq.~\ref{eq:d3dplummer}).

Eq.\ \ref{eq:jeans} can be used to fit directly  the  velocity dispersion profile estimated from the  observations (Fig.\ \ref{fig:compVdisp}).  This method has been used widely in the dynamical analysis of dwarf spheroidals, but it depends on the binning procedure (see Sec.\ \ref{sec:binned}) and it is not straightforward  to the take into account the $\vlos$ uncertainties of the single stars. In our analysis, we decided to exploit all the information of our Sculptor members catalogue (Sec.\ \ref{sec:memb}) considering the $\vlos$ of each star (see e.g.\ \citealt{pascale18}).

In particular, the posterior distribution for the DM halo parameter $\vec{\theta}=(r_\te{c}, M_\te{200})$   are estimated exploiting the Bayes theorem and  using the (logarithmic) likelihood
\begin{equation}
\ln \mathcal{L} = -0.5  \sum^{N_\te{Scl}}_{i=1} \left( \frac{\left(V_\te{los,i} - V_\te{sys}\right)^2}{\delta V^2_\te{los,i} + \sigma^{2}_\te{los}(R_i; \vec{\theta})} + \ln \left( \delta V^2_\te{los,i} + \sigma^{2}_\te{los}(R_i; \vec{\theta}\right) \right),
\label{eq:like}
\end{equation}
where $V_\te{los,i}$, $\delta V_\te{los,i}$ and $R_i$ represent the los velocity, its related error and the projected (circular) radius of the $i$-th star among the $N_\te{Scl}$ Sculptor's members. The systemic velocity $V_\te{sys}$ can be treated as a nuisance parameter. We have assumed prior probability distribution uniform for $r_\te{c}$ ($P(r_\te{c}/\te{kpc})=\mathcal{U}(0,10)$), uniform for $\log M_\te{200}$ ($P(\log M_\te{200}/M_\odot)=\mathcal{U}(7,12)$, corresponding to $P(M_\te{200})\propto M^{-1}_\te{200} $) and normal for $V_\te{sys}$ ($P(V_\te{sys}/ \kms)=\mathcal{N}(111.6,0.2)$, see Tab.\ \ref{tab:obsprop}). 

 The  likelihood in Eq.\ \ref{eq:like} derives from the assumption that the distribution of the los velocities are Gaussian.  This is not necessarily the case even in  simple spherical isotropic models (see e.g.\ \citealt{VDMvdisp}). Indeed, deviation from a normal distribution can be used as additional information to discern between different dynamical models \citep[see e.g.][]{Gervdis,breddelsdsph,pascale18}.
 We checked the $\vlos$  distribution of Sculptor's members in each bin used to estimate the Sculptor velocity dispersion profile (Sec.\ \ref{sec:binned}). We did not found any strong systematic deviation from a Gaussian in any of them. Therefore, we conclude that the assumption of normal distributed $V_\te{los}$ is good enough  for the purpose of this work.

Finally, the posterior distributions have been sampled using the  affine-invariant ensemble sampled Markov chain Monte Carlo method    implemented in the \texttt{Python} module \texttt{emcee} \citep{emcee}. We used 300 walkers evolved for 300 steps after 100 burn-in steps. 
% * <giuliano.iorio89@gmail.com> 2018-09-13T07:36:59.604Z:
%
% ^.

\subsection{Best-fitting dynamical models} \label{sec:dyn:results}

\begin{figure}
\centering
\centerline{\includegraphics[width=0.9\columnwidth]{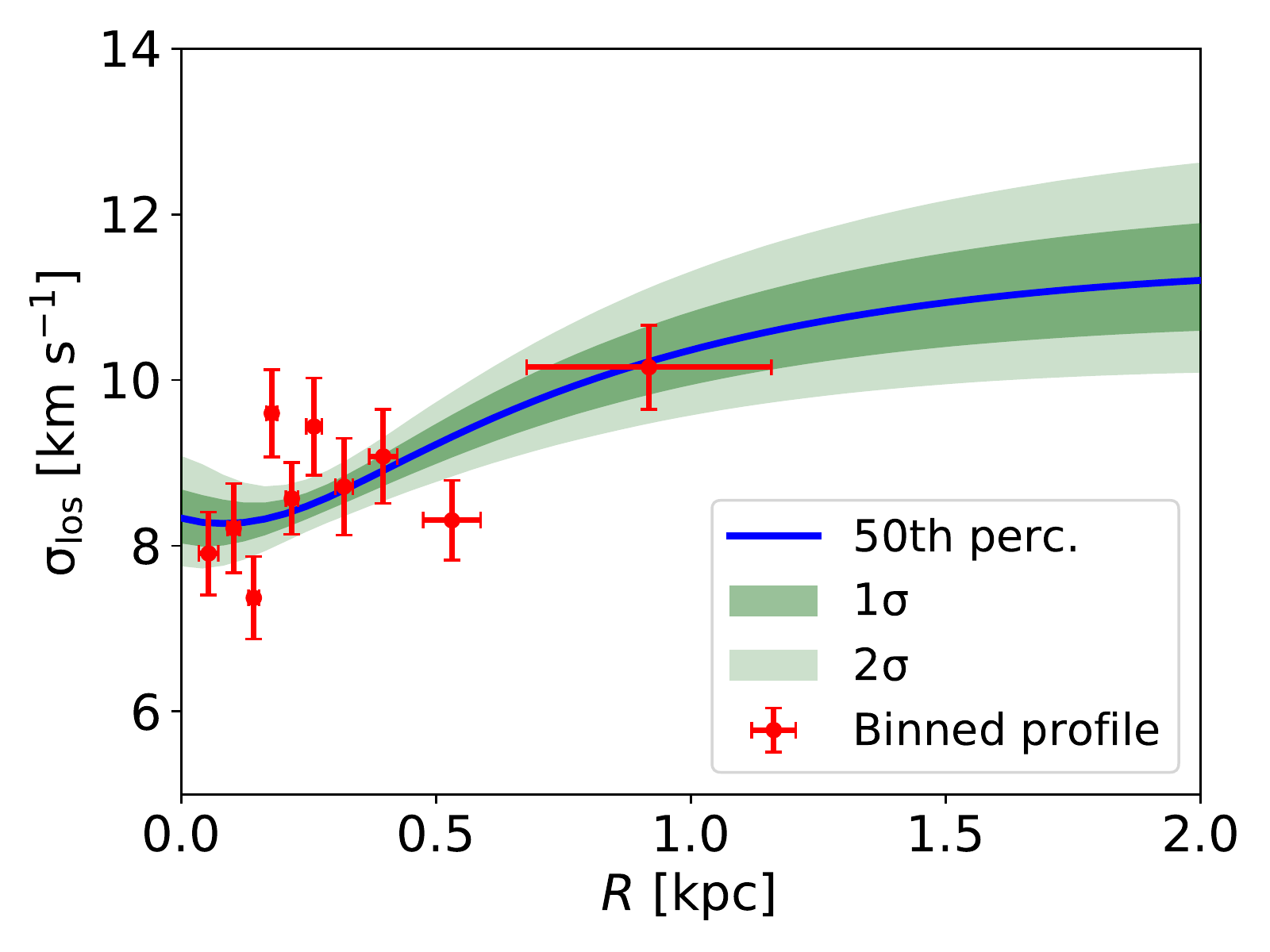}}
\caption[]{  Line-of-sight velocity dispersion profile of Sculptor (dots) compared with that of our dynamical model (curve and bands). In the model we show the posterior distribution for $\sigma_\te{los}(R)$ obtained using the likelihood in Eq.\ \ref{eq:like} and assuming $D=86$ kpc as  Sculptor distance (FH18 case, see Sec.\ \ref{sec:orbinv}).  The  line shows the median  $\sigma_\te{los}(R)$, while dark and light green bands indicate, respectively, the $68\%$ ($1\sigma$) and $95\%$ ($2\sigma$) confidence intervals. The red points are the same as in  Fig.\ \ref{fig:compVdisp}, assuming $D=86$ kpc. We stress that the  model does not fit the binned $\sigma_\te{los}$ profile but is determined using individually the line-of-sight velocities of Sculptor's members.}
\label{fig:dynbestfit}
\end{figure}

The posterior distributions of the predicted velocity dispersion profile is shown in Fig.\ \ref{fig:dynbestfit} for the FH18 case (see Sec.\ \ref{sec:orbinv}) compared with the observed velocity dispersion profile (Sec.\ \ref{sec:binned}). The results are similar also considering  the Sculptor distance assumed in the EH18 orbit.
The posterior distributions of  $\rc$ and $M_\te{200}$ indicate that these two parameters are quite correlated so that we can obtain similar results with a lower (higher) virial mass  and a smaller (larger) core radius.  However, in all the analysed cases the core radius is not compatible with 0, rather it is of the same order of the (projected) stellar half-light radius and it is about  $8\%$ of the NFW scale length $r_\te{s}$. 

The estimated core lengths are compatible with the theoretical predictions of transformation from primordial cuspy to cored DM density profiles due to effect of baryons (e.g.\ \citealt{nipoticore,Bermejo-Climent18}).
%In particular,  the formation of cores of the size of the half-light radius in \vir{classical} Milky Way dwarf spheroidals is compatible with estimates of the energy input by explosions of supernovae type II based on the observed star formation history of these galaxies \citep{Bermejo-Climent18}.
The presence of a core in the DM halo model is needed to reproduce the  velocity dispersion profile in the inner part of Sculptor (Fig.\ \ref{fig:dynbestfit}), at least under the assumption of isotropic stellar velocity distribution. 
%However, we cannot exclude that a similar could be produced assuming a tangential radial anisotropy of the stellar component \citep{tonry83}. 
In the case of cuspy DM profiles (like a pure NFW model), the galaxies result more resilient to the influence of the tidal fields \citep{penarrubia08, Pen10, frings17}.  In this context, the presence of extended cores in our model DM halo is compatible with the conservative strategy to set up our simulation in the most favourable conditions to maximise the effect of tides. For this reason, we did not run  simulations with \vir{cuspy} DM halos.

In the case of FH18 (assumed distance $D=86.0$ kpc) we obtained $V_\te{sys}=111.4\pm0.2 \kms$, $r_\te{c}=0.14\pm0.05$ kpc and $\log \left( \mvir / \msun \right) = 9.39\pm0.10$ ($c=14.89$), while for EH18 ($D=82.3$ kpc) we obtained $V_\te{sys}=111.4\pm0.2 \kms$, $r_\te{c}=0.15\pm0.04$ kpc and $\log \left( \mvir / \msun \right) =9.41\pm0.09$ ($c=14.83$).
%and considering the distance of EM18 ($D=92.4$ kpc) we obtained $r_\te{c}=0.21\pm0.05$ and $\log \mvir=9.31\pm0.10$. 
The systematic differences in the DM halo parameters are mostly due do the fact that, assuming different  distances for Sculptor, the physical distribution of Sculptor's members as well the physical scales (e.g. $\rc$ and $b_*$) are stretched or shrunk (EH18) with respect to the fiducial case (FH18). The DM dynamical model adapts to these changes reducing (or increasing) the core length and increasing (or reducing) the virial mass.
However, we verified that the mass within $r\approx 1\kpc$ (a radius containing most of Sculptor's members) is essentially independent of the  distance ($M_\te{DM}\sim10^8 \msun$, in agreement with the estimate by \citealt{zhu16}). 
Recently, \cite{errani18} estimated $M_\te{DM} = 3.16^{+0.82}_{-0.65} \times 10^7 \msun$ at radius $r=1.8 R_\te{h}\approx0.5 \ \kpc$ using a method that minimises the uncertainties due to the assumptions on the DM halo parameters. This value is largely consistent with our mass estimate at the same radius: 
$M_\te{DM} = 3.35\pm0.12 \times 10^7 \msun$  for FH18 and
$M_\te{DM} = 3.32\pm0.12 \times 10^7 \msun$ for EH18. 
We also estimated the slope of the mass profile as 
$\Gamma(r)={\rm d} \ln M_\te{tot}(r)/ {\rm d} \ln r$, finding, at the half-light radius $r=R_\te{h}$, $\Gamma=2.07 \pm 0.10$ for FH18 and $\Gamma=2.10 \pm 0.10$ for EH18. 
%These values are consistent with a cuspy-like radial profile ($\Gamma=2$). This is expected from Eq.\ \ref{eq:dNFWc} since at $R=R_\te{h}$ we are outside the region of the core ($R_\te{h}\sim2r_\te{c}$) where $\Gamma\sim3$, but still inside the inner part of the NFW  profile ($R_\te{h} \ll r_\te{s}$). 
Exploiting the presence of two distinct chemodynamical stellar populations in Sculptor (see Sec.\ \ref{sec:theScl} and Sec.\ \ref{sec:dicussion}) \cite{w11} used a non parametric method, based on the only assumption  $M(\approx R_\te{h}) \propto R_\te{h} \sigma^2_\te{los}$ (see also \citealt{errani18}), to  measure $\Gamma = 2.95^{+0.51}_{-0.39}$ at Sculptor's half-light radius.    
Investigating possible systematic effects behind this discrepancy, we found that the main cause is a slight difference of the \vir{average} measured  velocity dispersion. 
In fact, we estimated the los velocity dispersion of the two components  modifying the likelihood in  Eq.\ \ref{eq:like}  so that it includes two Gaussian components with (constant) velocity dispersion ($\sigma_1$, $\sigma_2$) and the relative component fractions are set using the best-fit radial profiles reported in \cite{w11}. 
%This method is consistent  with both the method used in \cite{w11} and the computation of the velocity dispersion highlighted in \cite{errani18}.
We found $\sigma_1=7.01 \pm 0.4 \ \kms$ and $\sigma_2=10.35 \pm 0.4 \ \kms$, while \cite{w11} measured $\sigma_1=6.4 \pm 0.4 \ \kms$ and $\sigma_2=11.6 \pm 0.6 \ \kms$. These differences  could be caused by the methods used to take into account  the Milky Way interlopers and/or of the outliers \citep[see also][]{strigari17}.   
 A detailed study of the internal dynamics and mass content of Sculptor, which is not the focus of this paper,  could be improved  through a complete analysis of the distribution of individual stellar velocities \citep[see][]{pascale18}.

\begin{comment}
repeated the dynamical fit using (i) two subcomponents for the stars (see Sec.\ \ref{sec:dicussion} for details), (ii) assuming a simple core profile for the DM ($\rho_\te{DM}\propto (1+r/r_\te{c})^{-2}$, (iii) assuming a constant velocity dispersion and estimating $\Gamma$ as in \cite{w11}, (iv) using only stars in our  catalogue that come from \cite{w09}. In all the cases, we obtain again a cuspy-like profile $\Gamma\approx2$. However, it is interesting to note that in (iii) we found velocity dispersions for the two stellar components, $\sigma_\te{los,1}=7.01 \pm 0.4 \ \kms$ and $\sigma_\te{los,2}=10.35 \pm 0.4 \ \kms$, that are slightly different with respect what found   ($\sigma_\te{los,1}=6.4 \pm 0.4  \ \kms$ and $\sigma_\te{los,2}=11.6 \pm 0.6  \ \kms$) by  \cite{w11}.  Using these values together with the half-light of the two populations estimated in \cite{w11} 
\end{comment}

In order to analyse the effect of the the baryonic mass-to-light ratio on the results of the dynamical fitting, we repeated the fit (for the FH18 case) adding the baryonic mass-to-light ratio in the V-band, $\Upsilon_*$, as an additional parameter  assuming the uniform prior $P(\Upsilon_*)=\mathcal{U}(0.01,100)$.
The final posterior distribution for $\Upsilon_*$ is quite broad and asymmetric with a peak at $\Upsilon_*\approx7$, a rapid decrease down to extreme values   ($\Upsilon_*\approx16$) and a more \vir{gentle} decrease toward lower values. 
In conclusion, the addition of $\Upsilon_*$ as free parameter increases the uncertainties on the values of the DM halo parameters, but it does not improve significantly the quality of the fit (similar value of the best likelihood). In particular, if we limit the analysis to realistic $\Upsilon_*$ values  (e.g. from few tenth to 4) the final maximum likelihoods are very similar. Hence, we decided to fix the mass-to-light ratio to $\Upsilon_*=2$ in the following $N$-body simulations (with the only exception of the mass-follows-light models of Sec.\   \ref{sec:mfl}).

%% file: include/Simulations/Simulations.tex
\section{$N$-body Simulations} \label{sec:Nbodysim}

In order to investigate the effects of the Milky Way tidal field on the structure and kinematics of Sculptor, we ran a set of $N$-body simulations putting  Sculptor-like objects in    observationally motivated orbits (Sec.\ \ref{sec:orbinv}) around the Milky Way (Sec.\ \ref{sec:orbital:mw}). Our work is analogous to the investigation of the effect of tides on Fornax by \cite{Fornax15}. Following that work, we iteratively  changed the initial conditions of the Sculptor-like objects to reproduce the observed position and the observed properties of Sculptor (Tab.\ \ref{tab:obsprop}) in the last snapshot of the simulations.
Finally, considering the known \vir{intrinsic} properties of the simulated objects, we investigated whether the observed kinematics of the stars is a robust tracer for the current dynamical state of Sculptor (in other words, whether Sculptor can be considered stationary) or the tides are introducing non-negligible biases.

\subsection{$N$-body realisations of Sculptor} \label{sec:nrel}
The initial conditions for each $N$-body realisation of Sculptor have been produced using the Python module \texttt{OpOpGadget}\footnote{https://github.com/iogiul/OpOpGadget} developed by G.\ Iorio. 
 The radii of the star and DM particles are randomly drawn to reproduce the input densities $\rho_{*,\te{Nb}}(r)$ and $\rho_{\te{DM},\te{Nb}}(r)$, respectively. 
Then, we  assign to each star the spherical angles $\phi$ (azimuthal angle) and $\theta$ (zenithal angle) picking randomly from  the uniform distributions  $P(\phi)=\mathcal{U}(0, 2\pi)$   and  $P(\cos \theta)=\mathcal{U}(-1, 1)$. In this way, we sample uniformly the sphere at each radius obtaining a spherical distribution of matter. Finally, the Cartesian coordinates are obtained as
$x=r\sin \theta \cos \phi$, $y=r\sin \theta \sin \phi$ and $z=r \cos \theta$.
In the next step, the velocity components of each star are assigned  through an acceptance-rejection technique assuming isotropy and using the two-component ergodic distribution functions obtained with  Eddington's inversion \citep{BT}.
Once the positions and velocities have been assigned to each particle, the position ($r_\te{COM}$) and the velocity ($V_\te{COM}$) of the centre of mass (COM) are estimated and the phase-space coordinates of the particles are translated to  correct for any \vir{residual} displacement from $r_\te{COM}=0 $ and  $V_\te{COM}=0 $.   The method produces spherical and isotropic $N$-body realisations of systems in equilibrium, as we verified by running simulations with the same initial conditions of those presented in this work, but in isolation, i.e.\ without the MW external potential.

We produced Sculptor-like objects using the stellar density law
\begin{equation}
\rho_{*,\te{Nb}}(r;M_*,b_*)=\rho_*(r;M_*,b_*)\exp\left[- \left(\frac{r}{r_\te{t}}\right)^2\right],
\label{eq:starNbody}
\end{equation}
where $\rho_*$ is the Plummer profile in Eq.\ \ref{eq:d3dplummer}, $M_*$  indicates the mass enclosed within the projected radius $R_*=1.5^\circ$ and $b_*$ is the Plummer core radius (see Sec.\ \ref{sec:dstar}).
Similarly, for the DM halo we used
\begin{equation}
\rho_{\te{DM},\te{Nb}}(r; \mvir, \rc)=\rho_\te{DM}(r; \mvir, \rc)\exp\left[- \left(\frac{r}{r_\te{t}}\right)^2\right],
\label{eq:DMNbody}
\end{equation}
where $\rho_\te{DM}$ is the cored NFW profile in Eq.\ \ref{eq:dNFWc} and the parameters $\mvir$ and $\rc$ indicate the virial mass and the halo core radius (see Sec.\ \ref{sec:dDM}), respectively.
The  truncation radius is needed because both the Plummer and the NFW profile have a non-negliglible probability to have particles up to infinity. In particular the NFW mass diverges when $r\rightarrow\infty$. Moreover, from a physical point of view, the DM halo of Sculptor is expected to be truncated due to the presence of the Milky Way.
We set $r_\te{t}=8.5\ \kpc$ for  FH18 and  $r_\te{t}=6.0\ \kpc$ for EH18, which, at the starting positions of the simulations, are of the order of (but slightly larger than) the Jacobi radius in the restricted three-body problem (the primary bodies are the MW and Sculptor, while the third body is a Sculptor's star, see \citealt{BT}).  
%In our iterative process the total mass of Sculptor changes at each simulation run, therefore we decided to use a conservative approach and set a constant $r_\te{t}$ assuming a slightly larger value than the estimated truncation radius considering the Sculptor parameters at the beginning of the iterative approach.  
%and $r_\te{t}=5.5\ \kpc$ for the EM18.  
The specific choice of the values of $r_\te{t}$ is not expected to be important for the final results of this work, because most of the DM halo particles close or beyond the truncation radius are stripped during the pericentric passages. Moreover, since the assumed truncation radii are much larger than the other involved physical scales ($b_*$, $\rc$), we are confident that the presence of a density truncation does not have any effect on the results of the dynamical fit in Sec.\ \ref{sec:dynmod}.
 
For each $N$-body realisation we used $N_*=5\times10^4$ stellar particles and $N_\te{DM}=4\times10^6$ DM particles. Since the total masses of the stellar and DM components change in the different analysed cases (FH18, EH18) and in different stages of the iterative process, the mass ratio between DM and stellar particles changes accordingly. It ranges between $\approx 2.5$ for the first run of EH18 and $\approx 8$ for the last run of EH18 (see Col.\ 8 in Tab.\ \ref{tab:result}). We re-run the simulations of the last iterative steps of FH18 and EH18 halving the number of star particles ($N_\te{*}=2.5\times10^4$) to halve the DM to stellar particle mass ratio.  The properties of the last snapshots of these additional simulations are compatible with the results presented in the following sections. Therefore we conclude that the results of this work (see Sec.\ \ref{sec:fresults}) are not influenced by the adopted particle mass ratios.

\subsection{Set up of the simulations}

\subsubsection{$N$-body code} \label{sec:orb}
The simulations have been performed using the collisionless code \texttt{FVFPS} \citep{code,Nipoti03}
with the addition of the \JP \MW model (see Sec.\ \ref{sec:orbital:mw}) as  external potential \citep{Fornax15}. 
We adopted  $\theta_\te{min}=0.5$ as the minimum value of the opening parameter and  we fixed the softening parameter to $\epsilon=10\ \te{pc}$.   We decided to use a constant time step,  $\Delta t= 0.03 t_\te{dyn}$, where 
\begin{equation}
t_\te{dyn} =  \sqrt{  \frac{1}{ G \bar{\rho}_\te{h} }  }
\end{equation}
is the initial dynamical time \citep{BT} of the Sculptor-like system and  $\bar{\rho}_\te{h}$ is the average 3D density estimated considering the total matter (stars and DM) content within the stellar half-mass radius $r_\te{h}$.
The dynamical time depends on the initial conditions of the $N$-body realisations, varying between $\approx 3 \times 10^5 \ \te{yr}$ and $\approx 5 \times 10^5 \ \te{yr}$ (see Col.\ 9 in Tab.\ \ref{tab:result}).

We do not include the gas in our simulations. However, we note that the star formation in Sculptor was already almost turned-off 8 Gyr ago (roughly the starting time of our simulations), hence it is likely that, in the analysed time window, the gas  was already absent or negligible with respect to the stellar and DM components. 

\subsubsection{Initial phase-space coordinates} \label{sec:icord}

We ran a series of simulations considering the two different orbital cases discussed in Sec.\ \ref{sec:orbinv} and summarised in Tab.\ \ref{tab:simdata}.
In each of the simulations, we set the initial phase-space coordinates of the $N$-body realisations 
at the first apocentre occurred within a lookback time of 8 Gyr.
 The FH18 represents the fiducial orbit obtained considering the  nominal proper motion reported in H18 and the nominal Milky Way parameters (see Sec.\ \ref{sec:orbital:mw}). The Sculptor-like object is placed at   $(x,y,z)=(-12.43,\ 98.05,\ -36.70) \ \te{kpc}$ with initial velocity $(V_x,V_y,V_z)=(-11.67,\ 44.19,\ 130.03) \ \kms$. The EH18 represents one of the most extreme (at $3$-$\sigma$ level) eccentric orbit compatible with the H18  proper motion measurements and with the errors on the Milky Way parameters.  The starting position is at  $(x,y,z)=(-8.21,\ 26.56,\ 90.33) \  \te{kpc}$ and  the initial velocity vector is $(V_x,V_y,V_z)=(13.34,\ -109.23,\  33.03)  \kms$.  
%The EM18 is the most extreme orbits analysed in this work in which the Sculptor-like object comes very close to the Galactic centre ($r_\te{peri}\sim18 \ \te{kpc}$) and it is based on the proper motion estimate by M18. 
%In this case, the starting position in the simulations is $(x,y,z)=(43.10,\ -60.16,\ -66.96) \ \te{kpc}$  and the initial velocity is $(V_x,V_y,V_z)=(-35.51,\ 45.08,\  -40.68) \ \kms$.

\subsubsection{Dynamical friction}
In the orbital investigation described in Sec.\ \ref{sec:orbital}, we have not considered dynamical friction \citep{BT}, which could affect the orbits of massive satellites, like dwarfs, reducing their pericentre, and  thus leading to stronger tidal effects.
We can have an estimate of  the dynamical friction time scale at a given Galactocentric radius, $r_\te{g}$, using equation (8.13) in \cite{BT}:
\begin{equation}
t_\te{f}(r_\te{g})=\frac{1.17}{\ln \Lambda} \frac{M_\te{MW}(<r_\te{g})}{M_\te{Scl}} t_\te{cross}(r_\te{g}),
\label{eq:tfric}
\end{equation}
where $M_\te{MW}(<r_\te{g})$ is the Milky Way mass within a sphere of radius $r_\te{g}$ and $M_\te{Scl}$ is the total mass of Sculptor estimated at the truncation radius (see Sec.\ \ref{sec:dyn:results} and Sec.\ \ref{sec:nrel}).
The crossing time, $t_\te{cross}$,  is defined as $r_\te{g}/v_\te{c}$, where $v_\te{c}=\sqrt{GM_\te{MW}(<r_\te{g})/r_\te{g}}$ is the circular speed at $r_\te{g}$. The Coulomb logarithm $\ln \Lambda=\ln b_\te{max}/b_\te{90}$ is estimated assuming   maximum impact parameter $b_\te{max}=r_\te{g}$ and   deflection radius $b_\te{90}$   equal to the half-mass (stars+DM) radius of Sculptor, $r_\te{h,tot}$.
At the  apocentres of the considered orbits ($r_\te{g} \approx 100\ \te{kpc}$, see Tab.\ \ref{ch3:tab:pmres})  $M_\te{MW} \approx 9 \times 10^{11} \ \msun$, $r_\te{t} \approx 6-12 \ \te{kpc}$, $M_\te{Scl} \approx  1-9\times10^9 \msun$ and $r_\te{h,tot} \approx 3-5 \ \te{kpc}$; while,  at the  pericentres  ($r_\te{g} \approx 40\ \te{kpc}$, see Tab.\ \ref{ch3:tab:pmres})  $M_\te{MW} \approx 4 \times 10^{11} \ \msun$, $r_\te{t} \approx 2-4 \ \te{kpc}$, $M_\te{Scl} \approx 0.3 - 2.0 \times10^ 9\  \msun$  and $r_\te{h,tot} \approx 1-3 \ \te{kpc}$.
The value of $t_\te{f}$  depends on the details of the considered family of simulations (FH18 or EH18), but we found that, considering both the apocentre and the pericentre, the dynamical friction time scale is always $t_\te{f}>13\ \te{Gyr}$. In conclusion, since the dynamical friction does not cause an appreciable difference in shaping the orbit of Sculptor, for simplicity we did not consider it in our simulations, in which the MW is represented by a fixed gravitational potential.

\subsection{Iterative method} \label{sec:iterative}

Following \cite{Fornax15} we have set up an iterative procedure in which we  tune the initial parameters of the $N$-body realisations (Sec.\ \ref{sec:nrel})  to reproduce in the last snapshot of the simulations the observed Sculptor properties, namely the position on the sky, the systemic velocity, the proper motions, the projected surface density, the los velocity dispersion profile and stellar mass  (see Tab.\ \ref{tab:obsprop}).  
The steps of the iterative procedure are as follows.

\begin{enumerate}
\item For the first simulation,  we make the $N$-body realisation assuming the stellar mass ($M_*=4.6\times10^6 \ \msun$ within $R_*=1.5^\circ$, see Sec.\ \ref{sec:dstar}) and the Plummer core length ($b_*=10.15\arcmin$)  estimated from the observations. Concerning the DM halo parameters $\rc$ and $\mvir$, we use the values estimated from the  dynamical models fitting  in Sec.\ \ref{sec:dyn:results}.
The initial position and the initial velocity are set as described in Sec.\ \ref{sec:icord}.
Accordingly, the initial integration time, $t_\te{sim}$, of each simulation  is set to reproduce the current Sculptor position and velocities in the last snapshot ($t_\te{sim}=6.9$ Gyr for FH18 and $t_\te{sim}=7.2$ Gyr for EH18, see Sec.\ \ref{sec:orbinv}). 
%$t_\te{sim}=8.0$ Gyr for EM18; see Sec.\ \ref{sec:orbinv}). 
However, they can change at different stages of the iterative process; see (iii) for further details.  
\item At the end of each simulation, we use the tools developed in \texttt{OpOpGadget} to \vir{simulate} an observation projecting the stellar particles on the plane of the sky (see Appendix \ref{app:mock}). We assume the centre of  the Sculptor realisation  as the position of  the centre of the system (COS)  of the stellar particle distribution (roughly, the peak of the stellar density) estimated with the iterative technique of \cite{Power03}.  We define  the COS velocity, as the mean velocity of the particle selected in the last iteration of the \cite{Power03} technique.
Then, we measure the stellar mass $M_{*,\te{sim}}$ within the projected radius $R_*=1.5^\circ$, the projected half-light radius $R_\te{h,sim}$  and  two values of the los velocity dispersion, $\sigma_\te{los,sim}(<R_\te{h,sim})$ and $\sigma_\te{los,sim}(<R_*)$, estimated considering all the stars within $R_\te{h,sim}$  and  $R_*$, respectively.
\item Each one of the observed parameters is sensitive mostly to a particular dynamical parameter, thus we update  our dynamical model using 
\begin{multline}
b_{*,\te{i+1}}=b_{*,\te{i}}  \frac{R_\te{h}}{R_\te{h,sim}}, \\ 
M_{*,\te{i+1}}=M_{*,\te{i}} \frac{M_*}{M_\te{*,sim}}, \\ 
r_{\te{c},\te{i+1}}=r_{\te{c},\te{i}}\frac{ \sigma_\te{los,sim}(<R_\te{h,sim})  }{\sigma_\te{los}(<R_\te{h})},\\
M_{\te{vir},\te{i+1}}=M_{\te{vir},\te{i}} \frac{ \sigma_\te{los}(<R_*)  }{\sigma_\te{los,sim}(<R_*)},\\
\label{eq:pcorr}
\end{multline}
where with the index $i$ we refer to the values used to make the currently analysed  $N$-body realisation.  
Due to the  gravitational effect of the DM particles lost in the tidal tails, it is not guaranteed that the COS of the stellar distribution matches perfectly the current position of Sculptor in the last snapshot.  In fact,  at the end of the simulations the COS position is still close to the predicted orbital track  (Sec.\ \ref{sec:orbital}), but ahead or behind with respect to the expected position. In other terms, this effect introduces a time offset ($\Delta t  \approx 10-300$ Myr) between the integration time predicted in the simple point-like orbit integration (Sec.\ \ref{sec:orbital}) and the integration time of the simulation, $t_\te{sim}$, needed to bring the simulated object in the  observed Sculptor position. 
We take into account this effect updating the integration time of the simulation as 
\begin{equation}
t_\te{sim,i+1}=t_\te{sim,i} - \Delta t.
\label{eq:tcorr}
\end{equation}

\end{enumerate}

The  steps (ii) and (iii) are repeated until the sum of the relative differences between the observed quantities %($R_\te{h}$,  $M_*(<R_*)$, $\sigma_\te{los}(<R_\te{h})$, $\sigma_\te{los}(<R_*)$) 
and those \vir{observed} in the last snapshot %($R_\te{h, sim}$,  $M_\te{*,sim}(<R_*)$, $\sigma_\te{los,sim}(<R_\te{h})$, $\sigma_\te{los,sim}(<R_*)$) 
of the simulation (see Eqs.\  \ref{eq:pcorr}) are no longer improved by the iterative procedure.
The simple time correction implemented with Eq.\ \ref{eq:tcorr} is able to tune the simulation integration time so that in the last snapshot of all  our final (last iteration) simulations the COS position and the COS velocity nicely match (relative errors within $5 \%$) the observed position, the systemic velocity and the proper motion of Sculptor (see Sec.\ \ref{sec:resit}).

%% file: include/Results/Results.tex
\section{Results} \label{sec:fresults}

\begin{table*}
\centering
    \tabcolsep 3pt
    \small
\begin{tabular}{c|ccc|cccc|cccccc}
\hline
 & \begin{tabular}[c]{@{}c@{}}$M_*$\\ {[}$10^6 \ \te{M}_\odot${]}\\ (1)\end{tabular} & \begin{tabular}[c]{@{}c@{}}$b_*$\\ {[}kpc{]}\\ (2)\end{tabular} & \begin{tabular}[c]{@{}c@{}}$M_{*,\te{tot}}$\\ {[}$10^6 \ \te{M}_\odot${]}\\ (3)\end{tabular} & \begin{tabular}[c]{@{}c@{}}$M_\te{200}$\\ {[}$10^9 \ \te{M}_\odot${]}\\ (4)\end{tabular} & \begin{tabular}[c]{@{}c@{}}$r_\te{c}$\\ {[}kpc{]}\\ (5)\end{tabular} &
 \begin{tabular}[c]{@{}c@{}}$c$\\ \\ (6)\end{tabular} &
 \begin{tabular}[c]{@{}c@{}}$M_{\te{DM},\te{tot}}$\\ {[}$10^9 \ \te{M}_\odot${]}\\ (7)\end{tabular} & \begin{tabular}[c]{@{}c@{}}$M_\te{tot}(<1\ \te{kpc})$\\ {[}$10^8 \ \te{M}_\odot${]}\\ (8)\end{tabular} & \begin{tabular}[c]{@{}c@{}}$f_\te{p}$\\ \\ (9)\end{tabular} & \begin{tabular}[c]{@{}c@{}}$t_\te{dyn}$\\ {[}Myr{]}\\ (10)\end{tabular} & \begin{tabular}[c]{@{}c@{}}$t_\te{sim}$\\ {[}Gyr{]}\\ (11)\end{tabular} & \begin{tabular}[c]{@{}c@{}}$f_*$\\$(<1\ \te{kpc}, <2\ \te{kpc})$ \\ (12)\end{tabular} & \begin{tabular}[c]{@{}c@{}}$f_\te{DM}$\\$(<1\ \te{kpc}, <2\ \te{kpc})$ \\ (13)\end{tabular} \\ \hline
\rowcolor[HTML]{C0C0C0} 
FH18$_{0}$ & 4.60 & 0.28 & 4.68 & 2.45 & 0.14 & 14.89 & 1.01 & 1.04 & 2.69  & 53.8 & 6.85 &  (96\%, 98\%) & (65\%, 55\%)   \\
FH18$_{14}$ & 4.62 & 0.26 & 4.69 & 6.31 & 0.13 & 13.80 & 1.95 & 1.49 & 5.20 & 44.69 & 6.79 & (97\%, 99\%) & (69\%, 60\%) \\
\rowcolor[HTML]{C0C0C0} 
EH18$_{0}$ & 4.60 & 0.27 & 4.68 & 2.57 & 0.15 & 14.83 & 0.84 & 1.05 & 2.23 & 52.41 & 7.19 & (89\%, 91\%) & (37\%, 23\%)  \\
EH18$_{17}$ & 4.66 & 0.22 & 4.71 & 22.39 & 0.12 & 12.38 & 2.94 & 2.23 & 7.81  & 35.76 & 7.07 & (95\%, 98\%) & (45\%, 35\%)  \\ \hline
\end{tabular}
\caption{Summary of simulation properties. For each analysed orbit (FH18 and EH18, see Tab.\ \ref{tab:simdata}), we report data for the first and the last simulations in the iterative cycle (Sec.\  \ref{sec:iterative}) as indicated by the subscript.   (1)(2) stellar mass within $R_*=1.5^\circ$ and core radius of the initial 3D stellar density profile (Eq.\ \ref{eq:d3dplummer} and Eq.\ \ref{eq:starNbody}); (3) total stellar mass  ($5\times10^4$ stellar particles); (4)(5)(6) viral mass, core radius  and concentration parameter of the initial 3D DM density profile (see Sec.\ \ref{sec:dDM}, Eq.\ \ref{eq:dNFWc} and Eq.\ \ref{eq:DMNbody}); (7) DM total mass ($4\times10^6$ DM particles);  (8)  total mass enclosed within $r=1$ kpc at $t=0$; (9) DM-to-stellar particle mass ratio; (10) initial dynamical time (it is used to set the simulation time step, see Sec.\ \ref{sec:orb}); (11) simulation integration time;  (12)  fraction of the initial stellar mass retained within  (1, 2) kpc in the last snapshot ($t=t_\te{sim}$);   (13) same as Col. (12), but for the DM mass.
}
\label{tab:result}
\end{table*}

In this Section, we present the results of the the families of simulations  FH18 and EH18 whose properties and initial set-ups are described in Sec.\ \ref{sec:orbinv}  and Sec.\ \ref{sec:Nbodysim}.  A subscript in the simulation name indicates what stage of the iterative cycle (Sec.\ \ref{sec:iterative}) we are referring to.
The index 0  indicates the initial  runs performed using the parameters obtained from the observations (see Tab.\ \ref{tab:simdata} and Sec.\ \ref{sec:dyn:results}; e.g.\ FH18$_{0}$ indicates the first simulations considering the FH18 orbit). 

\subsection{Results of the iterative process} \label{sec:resit}

\begin{figure*}
\centering
\centerline{\includegraphics[width=1.0\textwidth]{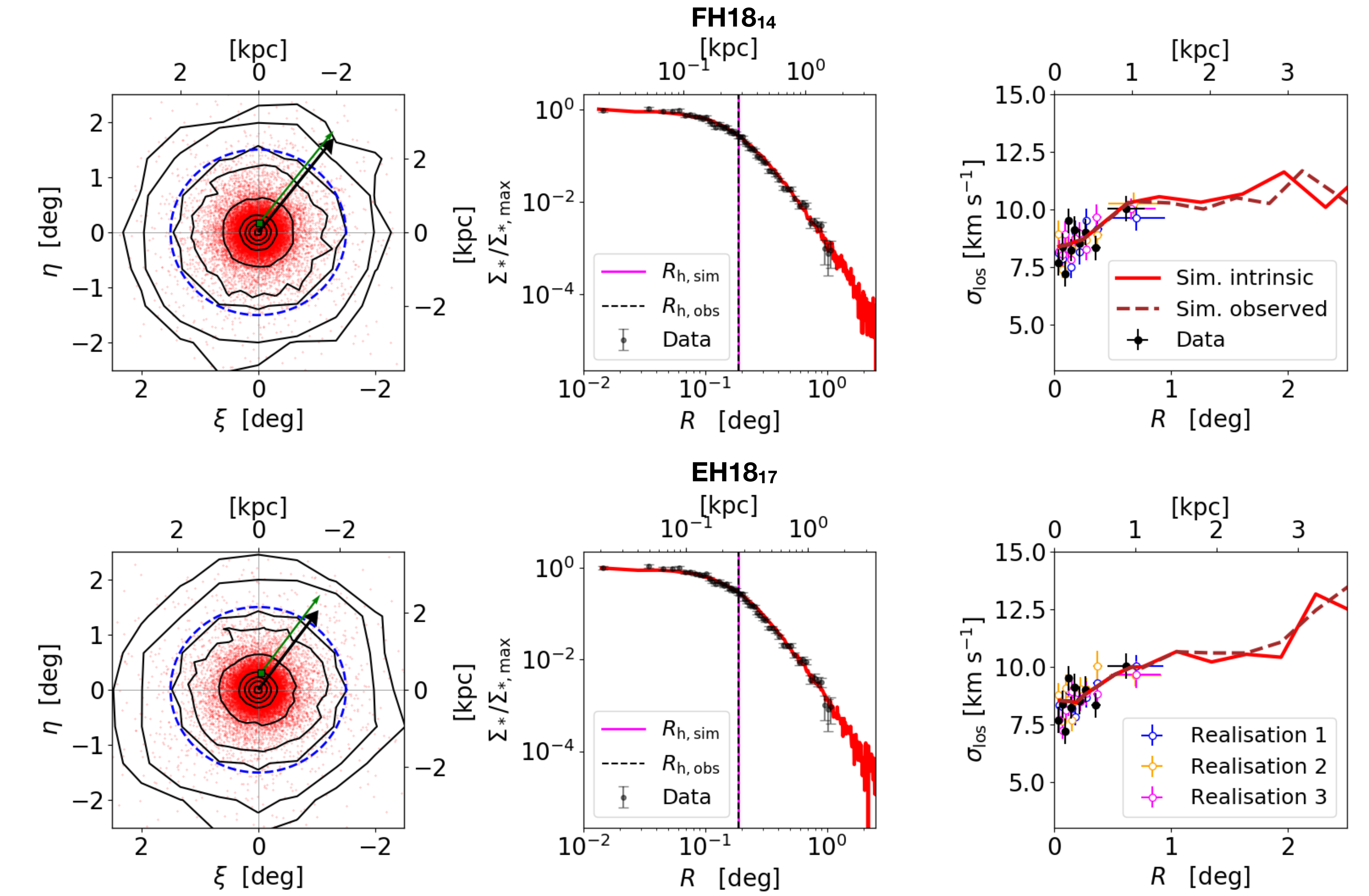}}
\caption[]{\vir{Observed} properties (see Appendix A) of the last snapshot of the $\te{FH18}_\te{14}$ (upper panels) and  $\te{EH18}_\te{17}$ (lower panels)  simulations. Left-hand panels: distribution of the stellar particles in the sky-projected equatorial $\xi-\eta$ plane (see  Fig.\ \ref{fig:smemb}); the dashed circle with radius $R_{*}=1.5^\circ$  is centred on the projected  COS (see Sec.\ \ref{sec:iterative}); the green square indicates the observed position of Sculptor with respect to the COS; the arrows indicate the direction of   Sculptor's proper motion (green-thin line) and of the  proper motion of the COS (black-thick line). Both proper motions have been corrected for the solar motions. The length of the arrows is  proportional to the proper motion magnitude. The black iso-density contours indicate regions where the surface density is the (0.005, 0.01, 0.05, 0.1, 1, 10, 25, 50, 90) percent  of the maximum value.  Middle panels: \vir{observed} surface density profile of the stellar particles (red line) compared with the (rescaled) Sculptor data (see Sec.\ \ref{sec:obs}); the vertical lines indicate the half-light radius measured for Sculptor (black-dashed) and estimated from the stellar particles (magenta), the two radii differ by less than 0.01 \%. Right-hand panels: \vir{observed} (brown-dashed line) and intrinsic (i.e. corrected for the Sun motion, red line) $\sigma_\te{los}$ profile  of the stellar particles compared with the binned $\sigma_\te{los}$ profile of Sculptor (black points); the blue, yellow and magenta points represent   binned profiles estimated with sub-samples of stellar particles (see Sec.\ \ref{sec:mockfit})  using the technique  described in Sec.\ \ref{sec:binned}  }
\label{fig:obs_combined}
\end{figure*}

The convergence of the iterative processes  (Sec.\ \ref{sec:iterative}) has been reached after 14 iterative steps  considering the fiducial orbit FH18 and after 17 steps for the more eccentric EH18 orbit. 
Fig.\ \ref{fig:obs_combined}  shows that, overall, the \vir{observed} properties of the last snapshots of the simulations  FH18$_{14}$ and  EH18$_{17}$ nicely agree with the observed properties of Sculptor. The final   positions of the simulated objects  are slightly off-set with respect to the current Sculptor position, especially in the EH18$_{17}$ simulation ($\approx 30\arcmin$),  due to the mismatch between the simulation orbit and the single-particle orbit (Sec.\ \ref{sec:orbinv}), as discussed in Sec.\ \ref{sec:iterative}. However, the distribution of the stellar particles and the kinematic properties (proper motions, velocity dispersion profile) show an almost perfect agreement.

Within the radial extent of the Sculptor dataset ($R_*=1.5^\circ$) the iso-density contours of the simulated objects remain smooth and there are no signs of tails. Even beyond this radius, neither the distribution of stars nor the stellar kinematics shows evident sign of tidal disturbance. This is the case for both the fiducial orbit FH18 and the  more eccentric orbit EH18. 
Beyond $R \approx 2^\circ$ the $\sigma_\te{los}$ profiles are inflated due to   the combination of few stars escaping from the system and the small number of tracers that increases the amplitude of statistical oscillations (see the simulation videos, links at the end of  Sec.\ \ref{sec:evres}).

 %The lack of stellar tails are evident also in the spatial  distribution of the DM and stellar particles shown in  Figs.\ \ref{fig:orb_FH18} and \ref{fig:orb_EH18}. 
Most of the stellar particles remains bound during the orbit integrations: at the end of the simulations only few percent of the stellar mass is lost considering  a radius of 2 kpc as a boundary (see Tab.\ \ref{tab:result}).   
This can be easily understood since the minimum 
truncation radius (estimated as the Jacobi radius, see Sec.\ \ref{sec:nrel}) at the pericentre of the orbits ($r_\te{t} \approx 3 \ \kpc$) is about twice the radius containing the bulk ($\approx 99 \%$, see Fig.\ \ref{fig:MfitMtrue}) of the stellar mass. 
We conclude that, given the current position of Sculptor and the most recent estimate of its proper motions, it is very likely that it has not developed any strong stellar  tails due to the interaction with the Milky Way tidal field. As a consequence, the spectroscopic dataset of Sculptor, as the one used in this work (Sec.\ \ref{sec:vdisp_prof}),  should not be contaminated significantly by tidally stripped stars. 

At the end of iterative  procedure the initial  DM halo has $r_\te{c}=0.13$ kpc and $M_\te{200}=6.31 \times 10^9 \ \msun$ (total mass of $1.95 \times 10^9 \ \msun$ considering the truncation) for FH18$_{14}$ and $r_\te{c}=0.12$ kpc and $M_\te{200}=2.24 \times 10^{10} \ \msun$ (total mass of $2.94 \times 10^9 \ \msun$ considering the truncation) for EH18$_{17}$ (see Tab.\ \ref{tab:result}).

\subsection{The evolution of the stellar and DM components } \label{sec:evres}

\begin{figure*}
\centering
\centerline{\includegraphics[width=0.9\textwidth]{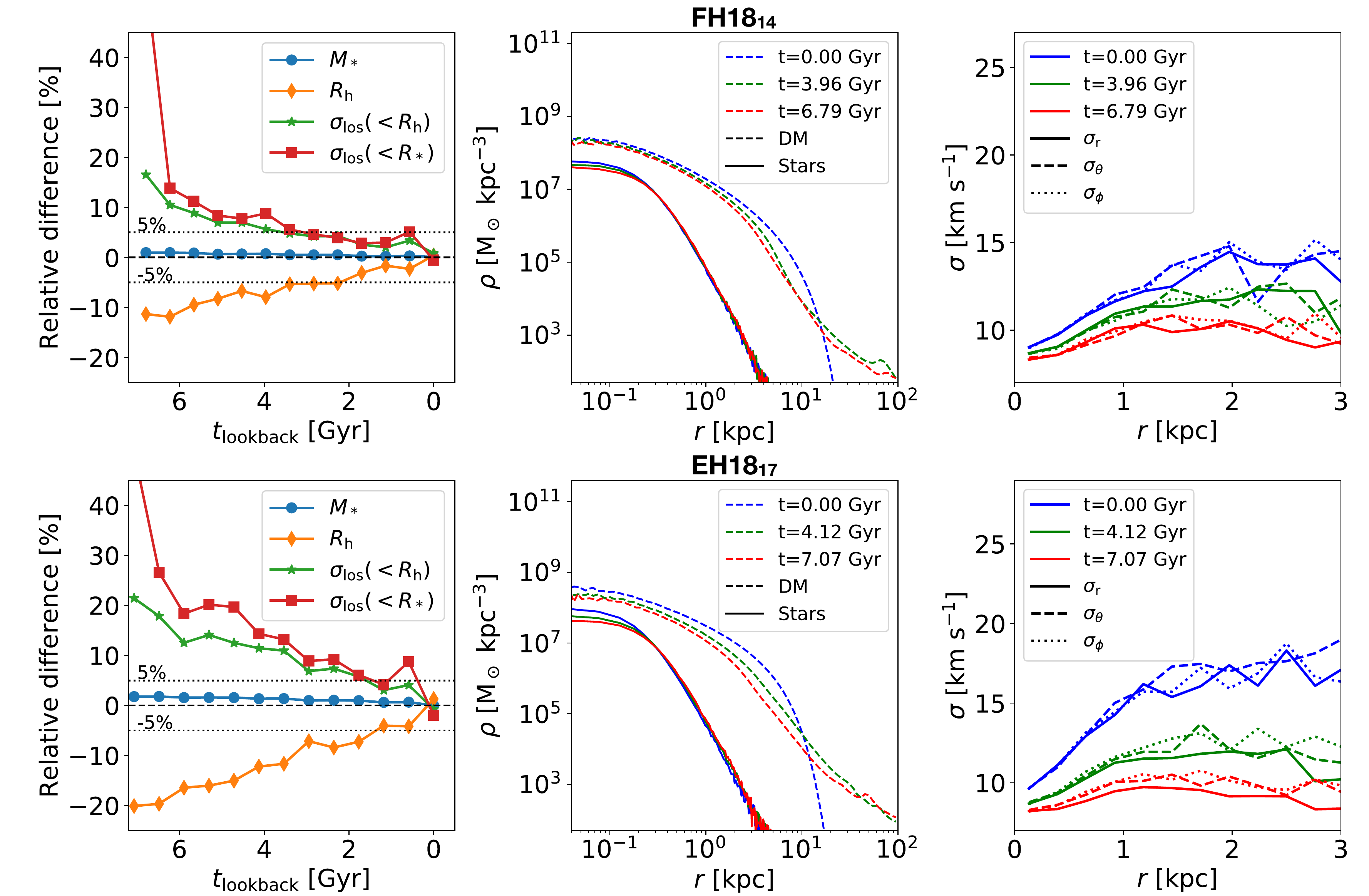}}
\caption[]{Time evolution of the properties of the FH18$_{14}$ (upper panels) and EH18$_{17}$ (lower panels) simulations. Left-hand panels: relative difference between the \vir{observed} properties (see Appendix \ref{tab:obsprop}) of the stellar particles in the simulation and the related properties observed in Sculptor (see Tab.\ \ref{tab:obsprop}): stellar mass within $R_*=2.25 \ \te{kpc}$ for FH18$_{14}$ and  $R_*=2.15 \ \te{kpc}$ for EH18$_{17}$ (blue circles), half-light radius $R_\te{h}$  (orange diamonds), los velocity dispersion profile within $R_\te{h}$ (green stars) and within $R_*$ (red squares); the value of $R_*$ is different for FH18$_{14}$ and EH18$_{17}$ because it is based on the definition of $R_*=1.5^\circ$ and the current assumed distance of Sculptor (see Sec.\ \ref{sec:memb} and Tab.\ \ref{tab:simdata});   see caption of Fig.\ \ref{fig:orb_fiducial} for the definition of $t_\te{lookback}$. 
Middle panels: 3D density profiles of the DM (dashed lines) and stellar (solid lines) particles at different times (initial conditions, blue lines; last snapshot, red lines; intermediate time, green lines). Right-hand panels: radial profile of the velocity dispersion components in spherical coordinates centred on the COS defined in Sec.\ \ref{sec:iterative} ($\sigma_\te{r}$, solid lines; $\sigma_\theta$ dashed lines; $\sigma_\phi$ dotted lines) at different times (initial conditions, blue lines; last snapshot, red lines; intermediate time, green lines).}
\label{fig:evol}
\end{figure*}
During the orbital evolution, the extended DM halo is influenced by tides quite instantaneously developing significant trailing and leading tails.
%Figs.\ \ref{fig:orb_FH18} and \ref{fig:orb_EH18} show the spatial distribution of DM and stellar particles at  three different stages  of the orbital evolution for the simulations  FH18$_{14}$ and EH18$_{17}$. 
Fig.\ \ref{fig:evol} shows that the density profile of the DM in FH18$_{14}$ evolves mostly in the outer part due to tidal stripping,  while the strong tidal shocks experienced by the satellite in EH18$_{17}$ (due to the smaller pericentric radius ) cause also a significant depletion of DM mass in the innermost part. 
 At the end of the simulations the fraction  of initial DM mass retained within 2 kpc is  60\% in  FH18$_{14}$ and only $35 \%$  in EH18$_{17}$ (see Tab.\ \ref{tab:result}). 
 Most of the evolution of the DM halo happens at the pericentric passages where the halo is heated by tidal shocks and it expands increasing the core radius, meanwhile in outer parts, it loses mass due to tidal stripping.  
 Therefore,   in order to reproduce the current properties of Sculptor, we had to start our simulations with smaller DM core radius and a larger DM total mass (about 2 and 3 times larger for FH18 and EH18, respectively, see Tab.\ \ref{tab:result}).
 
Figs.\  \ref{fig:evol}  show cleary that, in both the FH18 and the EH18 simulations,  the stellar component is not strongly affected by  the MW tidal field.
In particular, the left-hand panels of Fig.\ \ref{fig:evol} show that  the mass evolution within $R\approx 2 \ \te{kpc}$ is almost negligible: only few stellar particles are stripped during the evolution ($\approx 1 \%$ for FH18$_{14}$ and $\approx 2 \%$ for EH18$_{17}$), so no stellar tidal signatures are visible in the last snapshots of the simulations (see Sec.\ \ref{sec:resit}).
However, the evolution of the DM halo and the subsequent variation of the gravitational potential force the stellar component to find a new equilibrium modifying both its structure and its kinematics \citep{penarrubia08,pen09,Fornax15}.  
The stellar distribution tends to expand: in particular, the half-light radius increases by about 10\% in  FH18$_{14}$ and about 20\% in EH18$_{17}$ as shown in the left-hand panel of Fig.\ \ref{fig:evol}. 
The DM loss and the evolution of the DM density distribution causes a decrease of all the velocity dispersion components  (right-hand panels of Fig.\ \ref{fig:evol}). As a consequence,  the los velocity dispersion  decreases both in the inner parts (within the half-light radius) and in the outer parts (within $R\approx 2 \ \te{kpc}$) by about 15-20 \% in FH18$_{14}$ and about 25-30 \% in EH18$_{17}$ considering the last $\approx 6$ Gyr of orbital evolution.

We analysed also the evolution of the shape and of the anisotropy of the stellar component. During the orbital evolution the stellar axial ratios remain close to the initial value of 1,  more precisely they oscillate between 0.93 and 1 both in FH18$_{14}$ and  EH18$_{17}$.  Therefore, the flattening observed in Sculptor is likely an intrinsic property of  Sculptor, rather than induced by tides (but see \citealt{SandersCrater} for further details on the shape evolution of satellites in the MW tidal field).
Concerning the stellar velocity anisotropy, the right-hand panels of Fig.\ \ref{fig:evol} show that the values of the three spherical velocity dispersion components in the Sculptor frame of reference, $\sigma_\te{r}$, $\sigma_\theta$ and $\sigma_\phi$,  remain  close to each other for FH18$_{14}$ (i.e.\ the system remains isotropic) both in the intermediate and in the last phase of the orbital evolution.  Concerning EH18$_{17}$, in the last phases of the orbital evolution, the velocity dispersion along the radial direction tends to be slightly lower with respect to the tangential components resulting in a slightly tangentially anisotropic system. This is expected since the few stars that are lost in tails are preferentially the ones in radial orbits \citep{takashi2000,baumgardt2003}.  
Both in FH18 and EH18 The differences between the three components become larger at large radii ($r>2 \ \te{kpc}$), but this is mainly due to the increase of the Poisson noise: the same trend is apparent at $t=0 \ \Gyr$, when the system is isotropic by construction (see Sec.\ \ref{sec:nrel}).

Overall the main properties of the stellar component of the simulated Sculptor-like objects have changed very little (<10\%) in the last 3-4 Gyr (see left-hand panel of Fig.\ \ref{fig:evol}). 
Therefore, we conclude that  Sculptor can be reasonably well modelled as an isolated system in equilibrium.
 
 Videos of the evolution of the simulations FH18$_{14}$ and EH18$_{17}$ can be found online at \url{https://doi.org/10.5446/40807} and \url{https://doi.org/10.5446/40808}, respectively.

\subsection{Dynamical fit to the final snapshot} \label{sec:mockfit}
\begin{figure*}
\centering
\centerline{\includegraphics[width=1.0\textwidth]{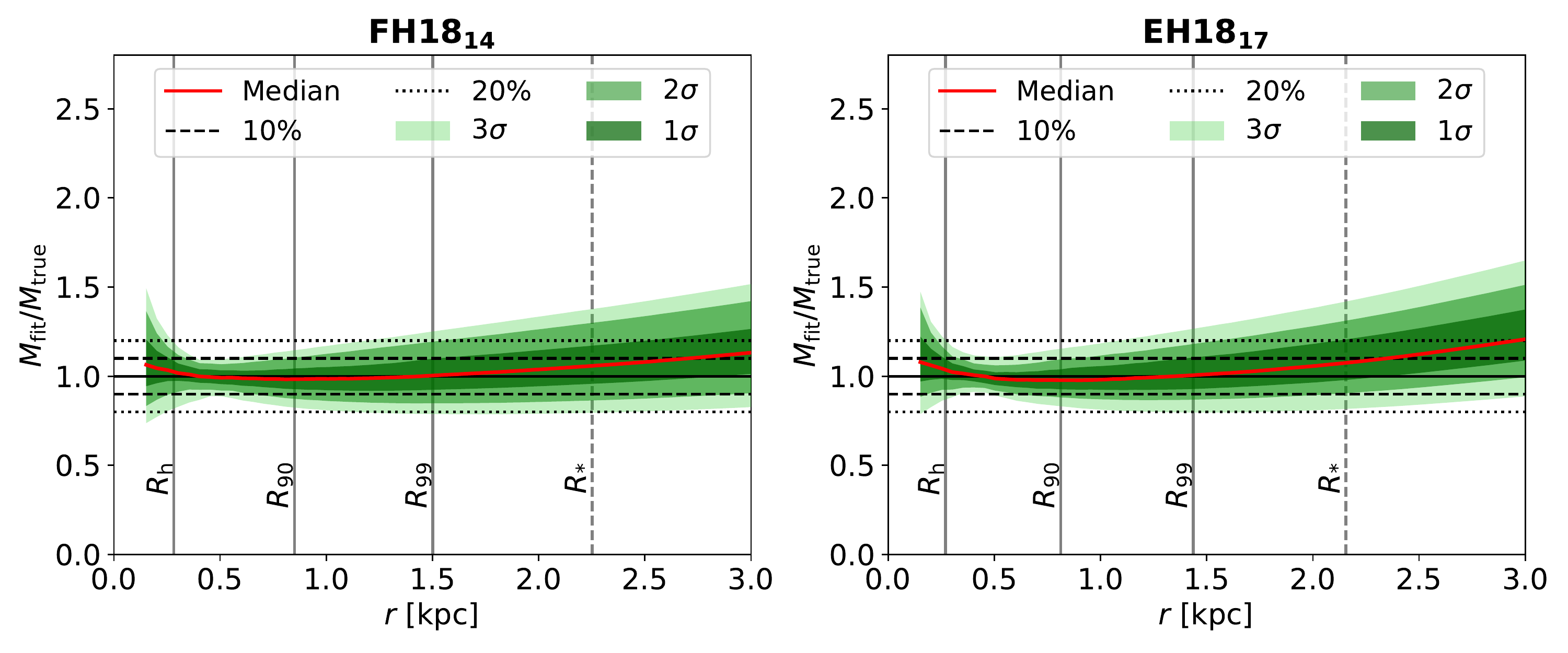}}
\caption[]{Ratio between the mass profile estimated from the dynamical fit and the true mass distribution in the last snapshot of the simulations FH18$_{14}$ (left-hand panel) and EH18$_{17}$ (right-hand panel). For each simulation we applied the dynamical fit procedure described in Sec.\ \ref{sec:dynfit} to  300 sub-samples of stellar particles selected to match the properties of our Sculptor dataset (Sec.\ \ref{sec:memb}, see text for further details). The red curve (median) and the green bands ($1\sigma$, $2\sigma$,  \textbf{$3\sigma$} levels) show the posterior distribution of the ratio between the mass profiles estimated using the 300 best-fit DM halo parameters and the true total mass distribution. The dashed and dotted horizontal lines highlight the regions where the mass differences are lower than 10\% and 20\%, respectively. The grey vertical lines indicate the  median  (over the 300 sub-samples)  of the projected  radii containing the $50\%$ ($R_\te{h}$), the $90\%$ ($R_\te{90}$) and the $99\%$ ($R_\te{99}$) of the stellar particles in the sub-samples.
The grey vertical-dashed line indicate the projected $R_*=1.5^\circ$ that represents roughly the radial extent of our dataset (see Sec.\ \ref{sec:dstar}). } 
\label{fig:MfitMtrue}
\end{figure*}

As an additional test for the dynamical state of Sculptor, we repeated the dynamical modelling applied to the data and described in Sec.\ \ref{sec:dynfit} to the stellar particles in the last snapshot of the  FH18$_{14}$ and EH18$_{17}$ simulations. 
In order to take into account the uncertainty due to the observational errors and data sampling, we did not use the complete sample of stellar particles, but rather we obtained 300 sub-samples as follows. (i) We bin the stellar particles using the same bin edges applied to the data to estimate the velocity dispersion profile  (see Sec.\ \ref{sec:binned}); (ii) for each sub-sample we randomly select a number of stars in each of the ten bins to match the number of stars in the observational data (192 particles in the last bin and 150 in the others); (iii) For each star we add Gaussian noise to the particles $V_\te{los}$ picking randomly from the $V_\te{los}$ errors distribution of the observational data (mean $\approx 2.3\ \kms$ and standard deviation $\approx 2\ \kms$).

At the end of this procedure, we obtain  300 sub-samples with stellar particles distributed along the projected radius roughly as the stars in our observed sample (see Sec.\ \ref{sec:memb}) and with errors that match the error distribution of the data. The velocity dispersion profiles estimated from three of these sub-samples can be seen in right-hand panels of Fig.\ \ref{fig:obs_combined}. We notice that the incomplete sampling causes fluctuations on the binned velocity dispersion profile so that at given bin different realisations can differ more than 1$\sigma$, where $\sigma$ is the error estimated as for the data (see Sec.\ \ref{sec:vdisp_prof}). These large fluctuations are   often seen in binned profiles of dwarf galaxies \citep[e.g.\ ][]{w09}, therefore one should be careful to interpret them as real kinematic features.
Finally, we applied exactly the same fitting procedure described in Sec.\ \ref{sec:dynfit}  to each sub-sample. Given the best-fit halo parameters ($r_\te{c}$ and $M_\te{vir}$), using Eq.\ \ref{eq:d3dplummer} and Eq.\ \ref{eq:dNFWc}  we computed  300 (total)  radial profiles of the total mass $M_\te{fit}$.
Fig.\ \ref{fig:MfitMtrue} shows the comparison between $M_\te{fit}$ and $M_\te{true}$ estimated  using the radial distribution of the stellar and DM particles in the last snapshot of the FH18$_{14}$ and EH18$_{17}$ simulations. 
Overall the two values show a good agreement. In particular, in the regions well sampled by the data  (within $R_{90}\approx0.57^\circ \approx 0.8 \ \te{kpc}$) the median of the mass ratio is everywhere $\approx 1$ and the errors due to the incomplete sampling are within the $20 \%$  at $3\sigma$ level.
Fig.\ \ref{fig:MfitMtrue} shows that beyond $R_{99}\approx 1^\circ \approx 1.5 \ \te{kpc}$, where there are few tracers, the mass content tend to be only slightly overestimated (by $10-20\%$).
%, but this overestimate is negligible comopared to the large dark-to-luminous mass ratio of the simulated objects (see Sec.\ \ref{sec:resit}) and estimated in Sculptor and other dwarfs \citep[e.g.][]{walker07}.
This test confirms the previous results  (see Sec.\ \ref{sec:evres})  indicating that tidal effects are not significant at the present-day position of Sculptor and along orbits compatible with the most recent proper motion estimates (\HP, \FP) and implies that Sculptor can be reliably modelled assuming that it is isolated and stationary.

\subsection{Mass-follows-light models} \label{sec:mfl}

So far we have assumed a cosmologically motivated DM halo of the satellite, which is much more extended than the baryonic component. In this case, we demonstrated that the DM halo is able to \vir{shield} the baryonic part of the satellite from the tidal field of the  MW. 
It is possible that, in absence of an extended DM halo, the tidal force can be effective enough to produce a flat and high los velocity dispersion profile \citep{aaronson}. 
This kind of model has been used by \cite{munoz08}   to obtain a  good match to the observed properties of Carina dSph (but see \citealt{UralCarina} for an alternative model including an extended DM halo). 
To explore this idea for Sculptor, we performed an additional suite of simulations assuming that the \vir{mass follows light}. 
In these models the $N$-body realisation of the satellite is obtained as one-component isotropic system following the Plummer density profile in Eq.\ \ref{eq:starNbody}. We considered two cases, in one we assume a mass-to-light ratio $\Upsilon_*=2$ corresponding to  a mass within $R_*=1.5^\circ$ of $M_*=4.6 \times 10^6 \msun$, in the other we assume a mass-to-light ratio $\Upsilon_*=12$ corresponding to $M_*=2.76 \times 10^6 \msun$. 
In the first case we are considering a pure baryonic representation of Sculptor, while in the second case the high  $\Upsilon_*$ (not realistic for a pure baryonic component) is the value needed to reproduce the $\sigma_\te{los}\approx 8-9 \ \kms$ in the innermost part of system without an extended secondary component (see Sec.\ \ref{sec:dyn:results}). 
In order to maximise the influence of tides, we put the two models in the  eccentric EH18 orbit (Sec.\ \ref{sec:orbital}) and we applied the same iterative approach used for the two-component models (Sec.\ \ref{sec:resit}) and described in Sec.\ \ref{sec:iterative}. 
We found that it is not possible to obtain a perfect match considering together the total light (or baryonic mass) within $R_*=1.5^\circ$,  the surface density profile and the los velocity dispersion profile. 
In particular, if we add more mass from the beginning, the system becomes more resilient to the tides so it is easy to match the observed surface density profile at the end of the simulations, but at the same time, we end with too much mass inside $R_*=1.5^\circ$. 
Tuning the initial mass to have the right amount of mass inside $R_*$,  we could create a system in which the final surface density profile is distorted by a large amount of tidally stripped stars. For this reason, the iterative steps have been halted after three runs for both models.

\begin{figure*}
\centering
\centerline{\includegraphics[width=1.0\textwidth]{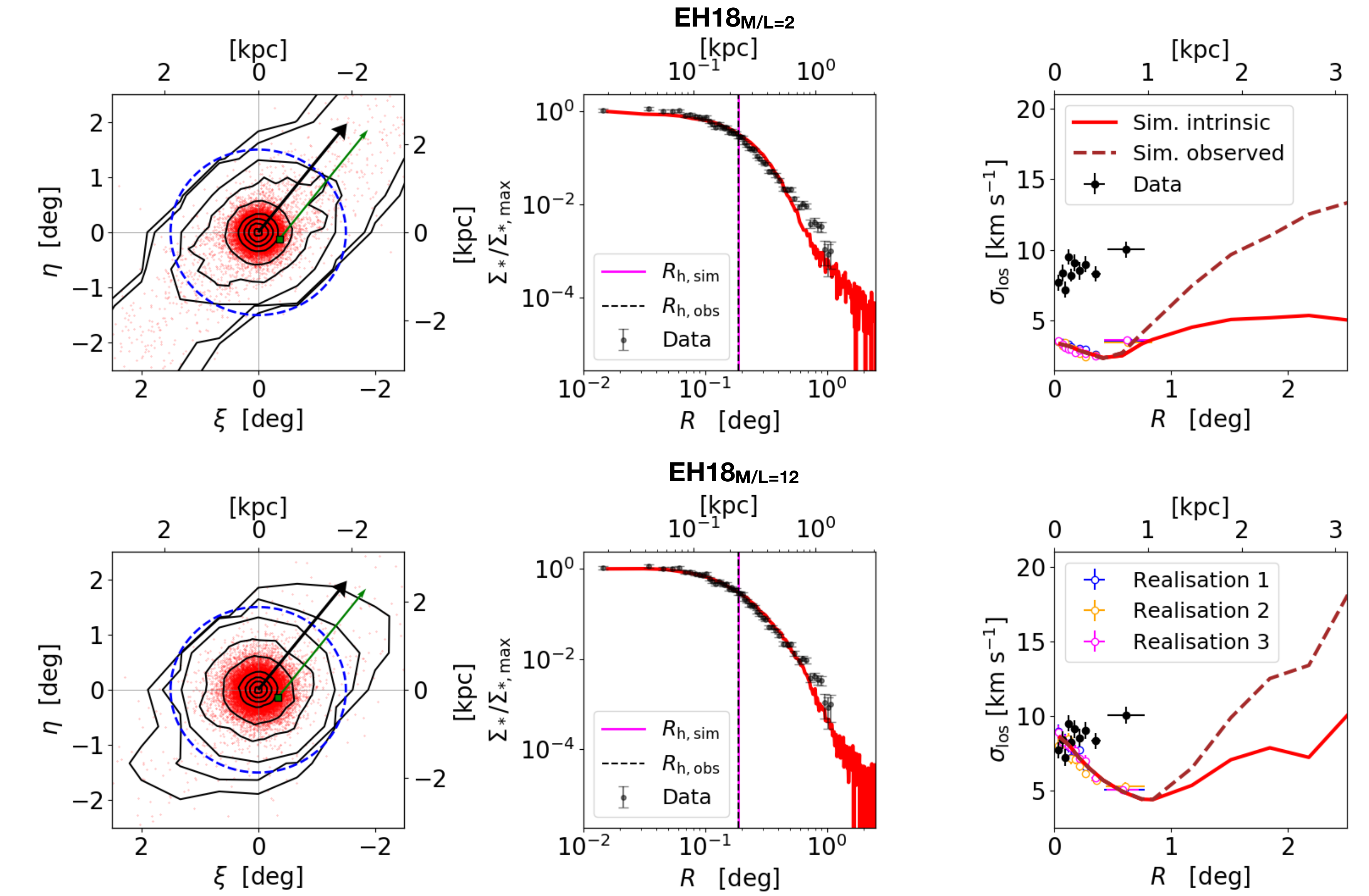}}
\caption[]{ Same as Fig.\ \ref{fig:obs_combined}, but for the last snapshot of the last-iteration simulation of the EH18 mass-follows-light model (see text) considering a  baryonic mass-to-light ratio $\Upsilon_*=2$ (upper panels) and $\Upsilon_*=12$ (lower panels).
 }
\label{fig:mfl}
\end{figure*}

The results of these simulations are shown in Fig.\ \ref{fig:mfl}. 
In both cases, the tidal stripping is effective enough to produce tidal tails that are particularly significant in the case with $\Upsilon_*=2$ (left-hand panels in Fig.\ \ref{fig:mfl}). 
In this model, the  tidal tails starts to effectively \vir{alter} the surface density profile around $R=0.6^\circ-0.7^\circ$ well inside the region covered by the photometric observations (Sec.\ \ref{sec:obs}). 
%As a consequence, it is not possible to perfectly reproduce the observed surface density profile, %although the half-light radius in the last snapshot perfectly matches the observed one (top-middle panel in Fig.\ \ref{fig:mfl}).
The mass loss is highly significant in the model with  $\Upsilon_*=2$  ($\approx 50\%$) and more moderate in the model  $\Upsilon_*=12$  ($\approx 10\%$). 
The innermost part is less affected by the tides, but the mass loss causes the half-light radius to shrink during the orbital evolution (about 20\% for $\Upsilon_*=2$ and about 5\% for $\Upsilon_*=12$).
The velocity dispersion profile remains similar to the initial one in the innermost part ($R\lesssim0.8^\circ$)  showing just a slightly decreasing trend with time (<10\%) in the case with  $\Upsilon_*=2$. 
Therefore, in the last snapshots, the los velocity dispersion profiles are still significantly decreasing with radius for both models and significantly lower than the Sculptor $\sigma_\te{los}$ for the model with $\Upsilon_*=2$.  Hence, neither model matches the los velocity dispersion profile of Sculptor (see right-hand panels in Fig.\ \ref{fig:mfl}). 
We also notice that the fact that the velocity dispersion is not enhanced in the inner parts means that the tidal tails are not aligned along the los \citep{klimentowsky07}.
At large radii ($R\gtrsim0.6-0.8^\circ$),  where the sample starts to be contaminated by stripped stars, the los velocity dispersion profiles of the simulations reverse the radial trend and start to increase with radius. However,  the $\sigma_\te{los}$ typical of  Sculptor ($\sigma_\te{los}\approx8-9 \ \kms$) are reached only in the very external regions ($R\approx2.5^\circ$) of the model with $\Upsilon_*=12$, far beyond the regions covered by  kinematic observations (right-hand panels in Fig.\ \ref{fig:mfl}). 

In conclusion, we found that, even in the absence of an extended halo component, the high-amplitude and flat observed velocity dispersion profile of Sculptor cannot be due to the effect of MW tides. Once again, this confirms that the Sculptor kinematic sample (Sec.\ \ref{sec:memb}) is practically free of contamination by tidally stripped stars.

A video of the evolution of the simulations with $\Upsilon_*=2$
can be found online at   \url{https://doi.org/10.5446/40809}.

%% file: include/Discussion/Discussion.tex
\section{Discussion and conclusions} \label{sec:dicussion}

In this work, we used ad-hoc $N$-body simulations to analyse the dynamical evolution of the Sculptor dSph in the tidal field of the MW.  
The dynamical models of the satellite and its backward orbits have been based on state-of-the-art observational data (see Sec.\ \ref{sec:theScl}).
Our final simulations are able to nicely reproduce the current position and velocities of Sculptor as well as the observed angle-averaged los surface density and velocity dispersion profile.  
The results of this work (Sec.\ \ref{sec:fresults})  indicate that, even considering one of the most eccentric orbit allowed by the data uncertainties  (primarily on the proper motions but also on other quantities; see Sec.\ \ref{sec:orbinv}), the direct effects of tides on the baryonic content of the satellite are negligible, and no significant tidal tails are developed. 
Even if not directly affected by tides, the stellar component of the satellite shows a certain degree of evolution because of the response to the significant mass loss experienced by the more extended DM halo. In particular,   the half-light radius increases and the  velocity dispersion decreases. 
However, in the last 2 Gyr the evolution proceeds very slowly, the structural and kinematic parameters of the baryonic component change less than about 5\% and the satellite in the last snapshots is close to equilibrium in its own gravitational potential. These results suggest that the assumption of dynamical equilibrium can be safely used to estimate the dynamical mass of Sculptor (Sec.\ \ref{sec:mockfit}). Moreover the lack of stellar tidal tails in the simulations indicates that we can consider the existing kinematic samples free from contamination by tidally stripped stars.
These results are similar to what obtained by \cite{Fornax15} for the Fornax dSph and they are compatible with the results of \cite{penarrubia09}.

%The lack of development of stellar tidal tails in both the FH18 and the EH18 simulations suggests that %the  Plummer-like profile observed in Sculptor is likely an intrinsic property of the system and it is %not caused by a  transformation of a primordial King profile as can be the case of objects heavily %affected by tides \cite{penarrubia09}.  

Recently \cite{hammer1} suggested that the gravitational acceleration due to the MW could heavily affect the los velocity dispersion of satellites.
In classical dSphs, such as Sculptor, this could explain the observed high los velocity dispersion \citep{hammer2}. Since this is a purely gravitational effect, its imprint should be present in our simulations. However, we found that in all our simulations, either including an extended DM halo (Sec.\ \ref{sec:evres}) or assuming a mass-follows-light model (Sec.\ \ref{sec:mfl}), the evolution of the los velocity dispersion profile within the observed range is driven by the mass loss of the DM halo and/or the property of the baryonic component. 
This is evident also in the mass-follows-light models:
%where the lack of an extended DM halo should make the effect of the MW acceleration postulated by \cite{hammer1} clearly evident.
the  right-hand panels of  Fig.\ \ref{fig:mfl} show that at the half-light radius the los velocity dispersion is totally determined by the distribution of matter inside the galaxy  (in both shape and magnitude).
%while the \cite{hammer1} model predicts a  higher\footnote{This value is obtained  using the \JP MW potential model in the  Eq.\ B16 of  \cite{hammer1}} $\sigma_\te{los} \approx 12 \ \kms$.
%Therefore, we conclude that the effect described by \cite{hammer1} is not important in Sculptor and the observed high-magnitude and flat los velocity dispersion profile is a genuine signature of the presence of an extended DM halo surrounding the baryonic component.

The simulations that successfully reproduce Sculptor's observables have  dark-to-luminous mass ratios 5.9 (FH18$_{14}$) and 5.4 (EH18$_{17}$)  within the projected stellar half-mass radius (283 pc in  FH18$_{14}$ and  271 pc in EH18$_{17}$)   and 62.1 (FH18$_{14}$) and  55.9 (EH18$_{17}$)  within the projected $R_*=1.5^\circ$ radius (2.25 kpc in  FH18$_{14}$ and 2.16 kpc in EH18$_{17}$).  The loss of DM is substantial: the initial mass of the satellite was at least two (FH18$_{14}$) and up to four (EH18$_{17}$) times higher 7 Gyr ago than today. 

The results of this work are based on a number of working assumptions. Most of them are simplifications that we tuned with the idea to maximise the possible tidal effects. 
First of all, we used a time-independent external potential considering a \vir{fully grown} MW for the entire $\approx 7 \ \te{Gyr}$ of the orbital evolution. The used potential is a simple analytic model \JP that represents a \vir{heavy} MW ($M_{200}>2\times10^{12} \ \msun$, see Sec.\ \ref{sec:orbital:mw}). 
Although this model does not represent the state of the art of the dynamical modelling of the MW and it does not take into account the hierarchical mass evolution of the Galaxy, it maximises the influence of tides on Sculptor. Therefore, our results are robust in the sense that if the tidal effects are not important in the analysed case it is likely that this holds even considering more sophisticated and time-dependent MW potential models. 

Concerning the assumption of the Sculptor's orbit, we relied on the robust estimate of the proper motions given by $Gaia$ (\HP, \FP). 
However the reported formal errors do not take into account the $Gaia$  systematics \citep{Lind18}. An estimate of the large scale systematic error can be obtained using Eq. 18 in \cite{Lind18}. For the $\approx 1^\circ$ degree area analysed in \HP we get  $\delta_{\mu,\te{sys}}\approx0.028$ mas yr$^{-1}$ (see Tab.\ \ref{ch3:tab:pmres}). 
In addition, we repeated  the \HP orbital investigation (Sec.\ \ref{sec:orb}) using the larger systematic error ($\delta_{\mu,\te{sys}}\approx0.066$ mas yr$^{-1}$) reported in \cite{Lind18} for small angular scales. Increasing the proper motion error, the distribution of the pericentre radius becomes wider and the $3 \sigma$ lower limit decreases to $\approx 23$ kpc, while the pericentre of our most eccentric analysed orbit (EH18, see Tab.\ \ref{tab:simdata}) is contained within $2 \sigma$. We applied our iterative method considering a new extreme orbit with $r_\te{peri}=23$ kpc and we found that in this case the tides are strong enough to influence significantly also the baryonic component of the dwarf. The final results are very similar to the mass-follows-light models analysed in Sec.\ \ref{sec:mfl}, the significant mass loss causes a strong decrease of the velocity dispersion ($\sigma_\te{los}\approx 5 \ \kms$) that does not match the observed kinematics. Although our method is able to explore only a limited portion of the dwarf initial conditions, our analysis points out that  considering a highly eccentric orbit ($e>0.6$) it is challenging  to reproduce  the observed Sculptor's properties.
It is also important to notice that in our analysis, we did not take into account the presence of the Large Magellanic Cloud (LMC). This massive object ($\approx 10^{11} \msun$, \citealt{erkal18}) could significantly perturb the MW satellites orbits.
\SP explored this effect finding that the LMC perturbation tends on one side to decrease the pericentric radius and on the other side to increase the orbital period. %Considering the tides, these two effects should balance (smaller pericentric radius, but fewer pericentric passages) and we do not expect a significant increase of tidal effects on Sculptor.  
However, even in the extreme case of a very massive MW ($M_{MW}=2 \times 10^{12} \ \msun$) and  very massive LMC ($M_{MW}=2.5 \times 10^{11} \ \msun$), the  pericentre found by \SP  ($\approx 40 \ \kpc$) is larger than what we have analysed with the EH18 orbit ($r_\te{peri}\approx 36 \ \kpc$, see Sec.\ \ref{sec:orbinv}). 
Therefore, considering the results of this work (Sec.\ \ref{sec:fresults}), it is unlikely that the LMC has produced any significant tidal effect on the baryonic component of Sculptor. However, in the future, it will be interesting to investigate further and in greater details the influence of the LMC on the dynamics and kinematics of Sculptor and other MW satellites. 

The dynamical models that we used for the dark matter halo of the simulated satellites rely on the present-day ($z=0$) concentration-mass relation (Sec.\ \ref{sec:dDM}). However,  the assumed simulation times ($t_\te{sim}\approx7$ Gyr, see Tab.\ \ref{tab:result}) correspond to an initial redshift  $z\approx0.8$. At that redshift, the concentration parameter expected for the halo mass range analysed in this work can be as low as 9.  However, most of this evolution is, in fact, a pseudo-evolution driven by the change of critical density of the Universe \citep{diemer19}. In practice,  at given $r_\te{c}$ and $M_{200}$ the density profiles  based on the $c_{200}$-$M_{200}$ relations at $z=0$ and $z=0.8$ are very similar. As an additional check, we repeated  the dynamical fitting of los velocity distribution (Sec.\ \ref{sec:dynfit}) and we ran the simulation EH18$_{17}$ (Sec.\ \ref{sec:Nbodysim}) assuming the $c_{200}$-$M_{200}$ relation at $z=0.8$ \citep{diemer19}. We found that the final results are largely consistent with what we found assuming $z=0$ (Sec.\ \ref{sec:fresults}).

Another fundamental assumption is that we consider only spherical and  isotropic models. In Sec.\ \ref{sec:dynmod}, we demonstrated that an isotropic model with a cored DM halo can give a reasonable fit to observed los velocity distributions of Sculptor. 
%Given the mass-anisotropy degeneracy, the same results can be obtained with a cuspy halo and a tangential anisotropy  \citep{tonry83}. 
Using a cored DM halo we maximised the effect of tides since the baryonic component is less \vir{shielded} from tides with respect to a cuspy halo profile \citep{penarrubia08, Pen10, frings17}. 
Concerning the shape of the baryonic distribution, we  assumed for simplicity a spherical distribution, in order to limit the number of degrees of freedoms.  However, in a work analogous to the present one, but for the Fornax dSphs, \cite{Fornax15}, 
found that the results do not change when a flattened model is used instead of a spherical one. Given the similarity of the method and of the results of the two works, we are confident that the inclusion of a more sophisticated flattened model does not affect the main conclusions of our work. 

In the present work, we do not attempt a detailed study of the internal dynamics of Sculptor and we make a few simplifying assumptions. In particular, when we build the dynamical model (Sec.\ \ref{sec:dynmod}), we assume a single stellar population and isotropic velocity distribution. The most restrictive assumption is the assumption of isotropy. Considering two stellar populations would be useful only if they 
were allowed to be anisotropic.
About the velocity distribution, our assumption is conservative: if we had assumed radial anisotropy (as indicated by previous works on the internal dynamics of Sculptor, see e.g.\ \citealt{battagliasculp,ma18}), the DM halo would have been more massive than in the isotropic case and thus the tidal effects even milder than found in our simulations.
However, we stress that the DM halo assumed in our model, though derived under simplifying assumptions, is not unrealistic. For instance, \cite{pascale19}, allowing for the presence of two anisotropic stellar components, find that Sculptor is well modelled with a DM halo very similar to ours.

The last key assumption is about the use of \vir{dry} simulations without a gaseous component. Since we analysed the satellite evolution over the last $\approx 7  \ \te{Gyr}$, the SFH of Sculptor justifies this assumption  \citep{SculptorSFR,Bermejo-Climent18}.  Although the use of more sophisticated hydrodynamic simulations could allow us to explore more realistic initial conditions for the satellite \citep{mayer06,Nichols11,Kazantzidis17,hammer2}, the results of the present work suggest that, in the presence of an extended DM halo, it is difficult that significant tidal perturbations develop in the stellar component of Sculptor.

In conclusion, we would like to stress again that the use of ad-hoc $N$-body simulations, such as those presented here, is a valuable tool to complement the dynamical analysis of MW satellites. 
Only in this way, it is possible to explore in details the possible perturbations produced by the MW tidal field and verify the validity of equilibrium models. 
%To date, this kind of approach has been used to analyse the dynamical status of Fornax dSphs \citep{Fornax15}, Carina dSph \citep{UralCarina}  and Sculptor dSph (this work).
%In the future, we aim to apply this method also for other objects like Draco dSph and Sextans dSph.  
The release of Gaia data and of the forthcoming large surveys make it possible to have robust proper motion estimates  not only for classical dSph satellites but also for ultrafaint and low surface brightness  dwarfs (e.g.\ Crater 2,  \citealt{Crater2,caldwell17}; or Antlia 2, \citealt{Antlia2}). The investigation of the properties of these objects with $N$-body simulations based on realistic orbits is  a fundamental step in the attempt to understand their content of DM and their formation and evolution (see e.g.\ \citealt{SandersCrater}).

%% file: include/appendix/mockobs.tex
\section{Mock observations} \label{app:mock}

In order to compare  the final snapshot of the simulations with the observations of Sculptor, we simulate the observation of the stellar particles projecting them on the sky plane. The procedure consist in the following steps. 
(i) We modify the positions of the particles from  the Galactocentric  frame of reference (see Sec.\ \ref{sec:orbital:mw}) to a right-handed heliocentric frame of reference. In practice,   we translate the original frame of reference by $R_\odot$ along the $x$-axis and then we switch the direction of the $x$-axis, so that it points towards the Galactic centre. 
(ii) Given the positions of the particles with respect to the Sun we estimate the sky coordinates and we set the central coordinates of our simulated object as the sky coordinate of its COS (see Sec.\ \ref{sec:iterative}). 
(iii) We add to the particle velocities the extra components $-\vec{v}_\odot$ due to the motion of the Sun in the inertial Galactic frame of reference (see Tab.\ \ref{tab:simdata}). 
(iv) Finally, for each stellar particle we apply a  rotoreflection (combination of a rotation  and a reflection in a plane perpendicular to one of the rotation axes) of the heliocentric frame of reference, so that the new $z$-axis is the line connecting the Sun with the stars, the $x$-axis points toward the Galactic South-Cap and the $y$-axis is aligned with the Galactic longitude:
\begin{equation}
\begin{bmatrix}
    V_b       &  V_\ell  & V_\te{los}
\end{bmatrix}
=
\begin{bmatrix}
    V_x       &  V_y & V_z
\end{bmatrix} \mathbfss{R}_\te{rot},
\end{equation}
where 
\begin{equation}
\mathbfss{R}_\te{rot}=
\begin{bmatrix}
    -\cos(\ell) \sin(b)      & -\sin(\ell) & \cos(\ell) \sin(b)  \\
    -\sin(\ell) \sin(b)       & \cos(\ell) & \sin(\ell) \cos(b) \\
    \cos(b)       & 0 & \sin(b)
\end{bmatrix}.
\end{equation}
In this new frame of reference, the velocity along the $z$-axis is the observed $V_\te{los}$, while the velocities along the $x$ and $y$ axes represent the velocities $V_b$, $V_\ell$ along the Galactic latitude and longitude, respectively. 
Once we have all the coordinates and the velocity in the Galactic sky coordinates, we use the classical geometrical transformation to obtain also the same information in the equatorial sky coordinate system \citep[][]{johnson87}.
After this procedure, we obtain the sky position of the centre of the simulated object as well as the observed properties (sky-position, radial velocity and proper motion)  for each stellar particle in the snapshot.